\documentclass[%
 aip,
sor,
 amsmath,amssymb,footinbib,
 reprint,%
]{revtex4-1}

\usepackage{graphicx}
\usepackage{dcolumn}
\usepackage{bm}

\usepackage[utf8]{inputenc}
\usepackage[T1]{fontenc}
\usepackage{mathptmx}
\usepackage{float}
\usepackage{caption}
\usepackage{subcaption}
\usepackage{color}
\usepackage{xcolor}
\usepackage{tikz}
\usepackage{pifont}
\usepackage{bbding}
\usepackage{hyperref}
\usepackage{bibentry}
\usetikzlibrary{shapes}

\usepackage{xcolor}
\usepackage{bm}
\usepackage{hyperref}
\usepackage{graphicx}
\usepackage{siunitx}
\usepackage{natbib}

\definecolor{Red}{rgb}{1,0.0,0.0}
\renewcommand{\vec}[1]{\bm{#1}}
\newcommand{\tsor}[1]{\bm{#1}}

\usepackage{amsmath}
\makeatletter
\newcommand{\overleftrightsmallarrow}{\mathpalette{\overarrowsmall@\leftrightarrowfill@}}
\newcommand{\overrightsmallarrow}{\mathpalette{\overarrowsmall@\rightarrowfill@}}
\newcommand{\overleftsmallarrow}{\mathpalette{\overarrowsmall@\leftarrowfill@}}
\newcommand{\overarrowsmall@}[3]{%
  \vbox{%
    \ialign{%
      ##\crcr
      #1{\smaller@style{#2}}\crcr
      \noalign{\nointerlineskip}%
      $\m@th\hfil#2#3\hfil$\crcr
    }%
  }%
}
\def\smaller@style#1{%
  \ifx#1\displaystyle\scriptstyle\else
    \ifx#1\textstyle\scriptstyle\else
      \scriptscriptstyle
    \fi
  \fi
}
\makeatother
\newcommand{\te}[1]{\overleftrightsmallarrow{#1}}

\makeatletter
\newcommand*{\rom}[1]{\expandafter\@slowromancap\romannumeral #1@}
\makeatother

\begin{document}
\newcommand{\markerone}{\raisebox{-0.1em}{\tikz{\node[draw,scale=0.7,circle,fill=none](){};}}}
\newcommand{\markertwo}{\raisebox{-0.1em}{\tikz{\node[draw,scale=0.7,regular polygon, regular polygon sides=4,fill=none](){};}}}
\newcommand{\markerthree}{\raisebox{-0.1em}{\tikz{\node[draw,scale=0.7,diamond,fill=none](){};}}}
\newcommand{\markerfour}{\raisebox{-0.1em}{\tikz{node[draw,scale=0.7,cross](){};}}}

\preprint{AIP/123-QED}

\title{Investigating the Nature of Discontinuous Shear Thickening: Beyond a Mean-Field Description}

\author{Jetin E. Thomas}
\affiliation{Martin Fisher School of Physics, Brandeis University, Waltham, MA 02454}
\author{Abhay Goyal}
\affiliation{Department of Physics, Institute for Soft Matter Synthesis and Metrology, Georgetown University,
Washington, District of Columbia 20057, USA}
\author{Deshpreet Singh Bedi}
\affiliation{Martin Fisher School of Physics, Brandeis University, Waltham, MA 02454}
\author{Abhinendra Singh}
\affiliation{Benjamin Levich Institute, CUNY City College of New York, New York, NY 10031, USA.}
\affiliation{Pritzker School of Molecular Engineering, University of Chicago, Chicago, Illinois 60637, USA}
\affiliation{James Franck Institute, University of Chicago, Chicago, Illinois 60637, USA}
\author{Emanuela Del Gado}
\affiliation{Department of Physics, Institute for Soft Matter Synthesis and Metrology, Georgetown University,
Washington, District of Columbia 20057, USA}
\author{Bulbul Chakraborty}
\email[]{bulbul@brandeis.edu}
\affiliation{Martin Fisher School of Physics, Brandeis University, Waltham, MA 02454}

\date{\today}
\begin{abstract}
Dense suspensions can undergo a dramatic increase in viscosity at a critical value of the shear stress. This phenomenon, termed discontinuous shear thickening (DST), has been attributed to an increase in the fraction of particle interactions becoming frictional with increasing shear stress, and a successful mean-field theory has been developed to explain various accompanying rheological properties. On a microscopic scale, however, conventional structural analysis measures such as the grain-position pair correlation function show no significant changes with the onset of DST, though recent work has shown that similar analysis in the dual space of contact forces does lead to marked changes at this transition. Furthermore, experimental results have suggested the existence of higher-order microscopic correlations and the importance of incorporating fluctuations away from a mean-field description.  To this end, we use a higher-order cluster analysis tool to study the force networks obtained from simulations of dense suspensions to construct an effective interaction potential in force space. We show that there are significant changes occurring in this potential as a function of density and stress close to DST.  We discuss the implications of these observations on an emergent field theory of the DST transition.
\end{abstract}
\maketitle

\section{\label{sec:intro} Introduction} 
Dense suspensions of 
grains in a fluid display an increase in viscosity $\eta = \sigma_{xy}/\dot{\gamma}$ (thickening) as the confining shear stress ($\sigma_{xy}$) or strain rate ($\dot{\gamma}$) are increased. At a critical, density dependent shear-rate $\dot{\gamma}$, the viscosity increases abruptly: a phenomenon termed Discontinuous Shear Thickening (DST). In stress-controlled protocols, $\eta \sim \sigma_{xy}$ marks the DST boundary~\cite{Mewis:2012zl,Brown:2014uq}. Experiments have also observed changes in other components of the stress tensor such as the first normal stress difference, $N_1 = \sigma_{xx}-\sigma_{yy}$ close to the DST  regime \cite{royer2016rheological}. A mean-field theory~\cite{wyart2014discontinuous,cates2014granulation},  based on an increase in the fraction of close interactions becoming frictional (rather than lubricated) with increasing shear stress, has been extremely successful at predicting the flow curves and the DST flow-state diagram in the space of packing fraction, $\phi$ and shear stress or strain rate~\cite{mari2014shear,singh2018constitutive}. The presence of frictional forces and the nature of the contacts between the grains has been intensively scrutinized and investigated~\cite{comtet2017pairwise,fernandez2013microscopic,monti2019effect}. The physical picture of lubricated layers between grains giving way to frictional contacts when the imposed $\sigma_{xy}$ exceeds a critical value set by a repulsive force \cite{wyart2014discontinuous} provides a consistent theory  of DST~\cite{singh2018constitutive}, shear jamming fronts \cite{han2017constitutive} and instabilities of the shear-thickened state \cite{hermes2016unsteady, chacko2019dynamic}. 

The link between the constraints that such forces generate at the grain level and the emerging flow properties at the level of the whole suspension, such as normal stress differences \cite{royer2016rheological,hsiao2017rheological,jamali2019alternative}, points to the existence of long ranged microscopic correlations that have been, so far, elusive to any structural analysis~\cite{mari2014shear}. Conventional measures such as the microscopic pair correlation function of the grain positions, in fact, do not exhibit pronounced changes accompanying DST. As a consequence, although several features relating to the flow of dense suspensions can be well explained within the mean-field theory~\cite{wyart2014discontinuous,cates2014granulation}, the nature of the microscopic correlations underlying this transition remains far from clear~\cite{mari2014shear}.  A recent theory~\cite{ref:1} has proposed that a representation of the shear thickening grains through their force networks, which emerges naturally from the collective force balance constraints under flow, can help better elucidate those correlations. In this ``force space'', one can identify a correlation function that exhibits significant changes in its anisotropy across the DST transition. These correlations reflect the collective behavior triggered by changes in the nature of the {\it contact forces}, which often arise due to small changes in grain positions difficult to detect in any positional correlations. An interesting, distinctive feature of DST is that the shear stress increases less rapidly than the mean normal stress, and hence their ratio, the  macroscopic friction coefficient, {\it decreases} as the fraction of frictional contacts {\it increases}. This, and direct visualizations~\cite{mari2014shear}, indicate that there are important changes in the network of frictional contacts that are not captured by scalar variables such as the fraction of frictional contacts. Remarkably, a statistical theory based on the observed correlations in force space, provides a semiquantitative description of the macroscopic friction coefficient and of the rheological changes accompanying DST~\cite{ref:1}. 

The force-space based statistical theory, however, has only been explored through a mean field approach, while several experiments have indicated that long-range spatial correlations close to DST must be accompanied by large, intermittent stress fluctuations ~\cite{rathee-pnas,manneville2018uncovering,rathee2019localized}. A theory in force space able to include fluctuations and predict their qualitative change close to DST, beyond mean field approximations, could therefore provide significant new insight into the origin and the nature of the transition. { The existing theory~\cite{ref:1} is difficult to generalize beyond mean field because the effective potential obtained directly from the correlations in force space is anisotropic and has a very strong clustering tendency, which makes it difficult to use it to sample the corresponding microstates in force space. In this paper, we outline a different procedure to (a) systematically construct a coarse-grained  effective potential, which describes the behavior at large scales in force-space, and (b) go beyond the mean field theory using numerical simulations of this effective potential. The idea is that the representation of the shear thickening suspension in force space can be mapped into a many-body, statistical mechanics description of points (akin to particles) interacting through the coarse-grained effective potential. As a consequence, numerical simulations that solve the corresponding many body equations of motion for the interacting points may naturally provide a beyond-mean-field description in force space. The coarse-grained potential we obtain here is {\it central} but, remarkably, its changes with the imposed stress and the density of the suspension in real space capture the observed changes in the anisotropy of the correlation functions in force space.} As in the earlier work~\cite{ref:1}, we start from the distributions of macroscopic quantities that are measured in particle based simulations { in real space} of model DST suspensions~\cite{mari2014shear,singh2018constitutive}. The simulation results illustrate the changing nature of fluctuations as DST is approached and provide evidence of non-trivial, highly correlated fluctuations. We use the microscopic information contained in the simulations, i.e. the local forces exchanged by the grains, to construct, through the constraints of local force balance, the representation of the shear thickening suspensions in terms of the underlying force network. To quantify the collective behavior of the changes detected in the force space representation, we apply a cluster analysis tool and show that two distinct clustering scales emerge as DST is approached, reflecting a scale separation of contact forces. From the representations of different flow-states, we construct { an effective interaction potential in force space through a coarse-graining procedure, which is justified by the scale separation observed in the cluster analysis. We find that changes occur in this potential as a function of packing fraction and stress close to DST. Following a possible analogy with equilibrium statistical mechanics, molecular dynamics simulations 
based on this effective potential suggest that the different regions of the shear-thickening flow-state diagram stem from qualitatively different underlying force-space state diagrams across the transition, whose differences can be traced back to the changes in the effective potential. We discuss the implications that the observed changes in the coarse-grained potential may have for the force space representation of the suspension and outline the emerging questions for future work on an effective field theory of the DST transition.}

This paper is organized as follows. In Section \ref{sec:simulation}, we present a brief summary of the shear-stress controlled microscopic simulations.  Section \ref{sec:simresults} presents distributions of the strain rate obtained from time series of the simulations data.  In Section \ref{sec:forcetiles}, we present a short description of the force-space representation. This is followed by the clustering analysis of particle patterns and force-space patterns in Section \ref{sec:DBSCAN}. Finally, in Section \ref{sec:potential}, we present our results for the effective interaction potential, and  discuss the implications of our statistical analysis. 

\section{\label{sec:simulation} Simulation methods}
We simulate a two-dimensional monolayer of non-Brownian spherical frictional particles
that are immersed in a Newtonian fluid under an imposed shear stress $\sigma_{xy}$.
This gives rise to a time-dependent  shear rate $\dot\gamma$~\citep{Mari_2015, seto2019shear} and velocity field $\vec{v} = \dot\gamma(t)\hat{\vec{v}}(\bm{x}) = \dot\gamma(t)(y, 0)$.
Lees-Edwards periodic boundary conditions are used with $N=2000$ particles in a unit cell.
Bidisperse particles of radii $a$ and $1.4a$ mixed at equal volume fractions are used to avoid ordering during flow~\cite{Mari_2014}.
In the simulation scheme presented here, the particles interact through near-field hydrodynamic interactions (lubrication), a short-ranged repulsive force,
and frictional contact forces. 
The simulation model used here has been shown to accurately reproduce 
the experimentally measured rheology for dense shear-thickening suspensions~\cite{Seto_2013a,Mari_2014}.
%

The motion is considered to be inertialess, so that the equation of motion reduces to force/torque balance between hydrodynamic ($\vec{F}_{\mathrm{H}}$),
repulsive ($\vec{F}_{\mathrm{R}}$), and contact ($\vec{F}_{\mathrm{C}}$) interactions,
\begin{equation}
\vec{0} = \vec{F}_{\mathrm{H}}(\vec{X},\vec{U}) + \vec{F}_{\mathrm{C}}(\vec{X}) + \vec{F}_{\mathrm{R}}(\vec{X}), 
\label{eq:force_balance}
\end{equation}
where $\vec{X}$ and $\vec{U}$ { denote the positions and velocities/angular velocities of all particles, respectively.}
The repulsive force $\vec{F}_{\mathrm{R}}$ is conservative in nature and can be determined based on the positions $\vec{X}$ of the particles.
On the other hand, calculation of the tangential component of the contact force $\vec{F}_{\mathrm{C}}$ is more involved as it also depends
on the contact deformation history.

The translational velocities are non-dimensionalized by $\dot{\gamma}a$ and the shear rate and rotation rates by $\dot{\gamma}$.  
The hydrodynamic force is the sum of forces arising from the drag due to the motion of the particle relative to the surrounding fluid and the resistance to deformation imposed by the flow:
\begin{equation}
\vec{F}_{\mathrm{H}}(\vec{X},\vec{U}) =
-\tsor{R}_{\mathrm{FU}}(\vec{X}) \cdot \bigl(\vec{U}-\dot\gamma\hat{\vec{U}}^{\infty} \bigr)
+ \dot\gamma\tsor{R}_{\mathrm{FE}}(\vec{X}):\hat{\tsor{E}}^{\infty}, \label{eq:hydro_force}
\end{equation}
where $\hat{\vec{U}}^{\infty} = (\hat{\vec{v}}(y_1), \dots, \hat{\vec{v}}(y_N), \hat{\vec{\omega}}(y_1), \dots, \hat{\vec{\omega}}(y_N))$,
and $\hat{\tsor{E}}^{\infty} = (\hat{\tsor{e}}(y_1), \dots, \hat{\tsor{e}}(y_N))$ { is the normalized strain rate tensor}.
$\tsor{R}_{\mathrm{FU}}$ and $\tsor{R}_{\mathrm{FE}}$ are position-dependent resistance tensors
and include the
``squeeze'', ``shear'' and ``pump'' modes of pairwise lubrication~\citep{Ball_1997}, as well as one-body Stokes drag.
Regularization of the resistance matrix is achieved by introducing a small cutoff length scale $\delta=10^{-3}$, typical for non-Brownian suspensions~\cite{Mari_2014}.
This regularization emulates the occurrence of contacts between particles due, for example, to surface roughness.
The lubrication force is upper-bounded, and negative interparticle gaps $l$ (i.e., particle overlaps) are allowed in our simulations. 

We use a stablizing repulsive force decaying exponentially with the interparticle gap $l \geq 0$ as $|\vec{F}_R|  = F_0 \exp(-l/\lambda)$,
with a characteristic Debye length $ \lambda$. This force represents an electrostatic double layer interaction between particles.
In the simulations presented in this study, we use $\lambda = 0.02a$.

To model contacts between particles -- which occur only when the shear force is large enough to overcome the repulsive force $F_0$ -- we use linear springs { with both normal and tangential components}, as is commonly done in soft-sphere DEM simulations for dry grains~\cite{Cundall_1979, Luding_2008}. { Note, however, that there is no dashpot in this model, since hydrodynamic resistance provides the source of energy dissipation.}
The { corresponding normal ($k_{n}$) and tangential ($k_{t}$)} spring constants used here satisfy  $k_{\rm t} = 0.5 k_{\rm n}$.
The tangential and normal components of the contact force
$F_C^{(ij)}$ between two particles satisfy Coulomb's friction
law, i.e.,  $|F_{C,t}^{(ij)}| \le \mu|F_{C,n}^{(ij)}|$ with $\mu$ being the interparticle friction coefficient.
In this study, we use $\mu=1$.
This value of friction coefficient $\mu$ is comparable to experimentally measured values~\cite{Fernandez_2013},
where $\mu$ is reported to be in the range $0.6$--$1.1$ for polymer brush-coated quartz particles of diameter
$2a \sim10$ $\mu \rm m$, while it is higher than the value of 0.5 reported by Comtet {\it et al.}~\cite{Comtet_2017}.
Some softness is allowed at the contact; we tune the spring stiffness for each $(\phi, \sigma_{xy})$
such that the maximum overlaps between particles do not exceed \SI{3}{\percent}
of the particle radius during the simulation, thereby staying close to  the rigid limit \cite{Mari_2014,Singh_2015,singh2019yielding}.

The equation of motion~\eqref{eq:force_balance} is solved under the
constant shear stress $\sigma_{xy}$ constraint.
At any time during the simulation, the shear stress in the suspension is given by 
\begin{equation}
\label{eq:stress}
\sigma_{xy} = \Sigma_{12} = \dot\gamma \eta_0\biggl(1+\frac{5}{2}\phi\biggr)
+ \dot\gamma \eta_{\mathrm{H}}+\sigma_{\mathrm{R}}+ \sigma_{\mathrm{C}}
\end{equation}
where $\eta_0$ is the viscosity of the suspending fluid,
$\eta_{\mathrm{H}} \dot{\gamma} = \dot{\gamma} V^{-1} \bigl\{ (\bm{R}_{\mathrm{SE}}
-\bm{R}_{\mathrm{SU}}\cdot\bm{R}_{\mathrm{FU}}^{-1}
\cdot\bm{R}_{\mathrm{FE}} ) :\hat{\bm{E}}^{\infty}\bigr\}_{12} $ is the contribution of 
hydrodynamic interactions to the stress, 
and
$\sigma_{\mathrm{R,C}} = V^{-1}\bigl\{ \bm{X}\bm{F}_{\mathrm{R,C}}
-\bm{R}_{\mathrm{SU}}\cdot\bm{R}_{\mathrm{FU}}^{-1}\cdot\bm{F}_{\mathrm{R,C}}
\bigr\}_{12}$, where 
$\bm{R}_{\mathrm{SU}}$
and $\bm{R}_{\mathrm{SE}}$ are position-dependent resistance matrices
giving the lubrication stresses from the particle velocities
and resistance to deformation, respectively~\citep{Jeffrey_1992,Mari_2014},
and $V$ is the volume of the simulation box. 
At a fixed shear stress $\sigma_{xy}$ the shear rate $\dot\gamma$ is the dependent variable
which is calculated at each time step using~\citep{Mari_2015}
\begin{equation}
\dot{\gamma} = \frac{\sigma_{xy} - \sigma_{\mathrm{R}}-
	\sigma_{\mathrm{C}}}{\eta_0\Bigl(1+2.5\phi \Bigr) + \eta_{\mathrm{H}}}.
\label{eq:strainrate}
\end{equation}
The full solution of the equation of motion \eqref{eq:force_balance}
under the constraint of constant fixed stress \eqref{eq:stress}
thus reduces to calculating the velocity~\citep{Mari_2015}
\begin{equation}
\bm{U}  = \dot{\gamma}\hat{\bm{U}}^{\infty}
+
\bm{R}_{\mathrm{FU}}^{-1}\cdot
\bigl(
\dot\gamma
\bm{R}_{\mathrm{FE}}:\hat{\bm{E}}^{\infty}
+ \bm{F}_{\mathrm{R}}
+ \bm{F}_{\mathrm{C}}
\bigr).
\end{equation}
From these velocities, we update the positions at each time step.

Lastly, the unit scales for strain rate is $\dot{\gamma}_0 \equiv F_{\rm 0}/{6\pi \eta_0 a^2}$ 
and $\sigma_0 \equiv \eta_0 \dot{\gamma}_0 = F_{\rm 0}/6\pi a^2$ for the stress.

\subsection{Flow curves and flow state diagram}
\begin{figure}[t!]
\includegraphics[width=0.45\textwidth]{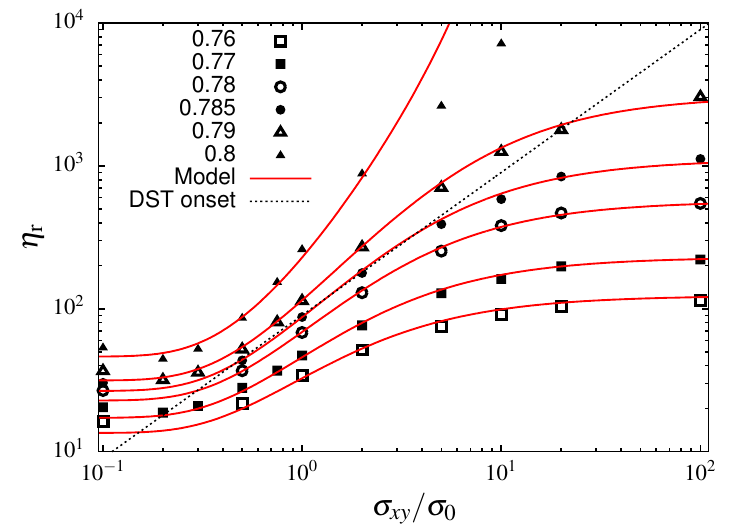}
\caption{
Relative viscosity $\eta_{\mathrm r}$ plotted as a function of
dimensionless applied stress $\sigma_{xy} / \sigma_0$. The symbols are simulation data for various packing fractions.
The solid lines are predictions from Eq.~\eqref{eq:eta_str_phi} for $\sigma^{\ast} = 0.78\sigma_{0}$. The dashed line represents $\eta_{\rm r} \propto \sigma_{xy}/\sigma_0$, representing DST.}
\label{fig:visc_str_model}
\end{figure}

Fig. ~\ref{fig:visc_str_model} shows relative viscosity $\eta_{\rm r}$ plotted as a function of scaled shear stress $\sigma_{xy}/\sigma_0$ for simulations at several packing fractions.
The relative viscosity data shows features typical of dense non-Brownian suspensions: shear thinning at low stress (arising due to the specific simulation model used here), thickening at intermediate stresses, and plateauing at high stresses $\sigma_{xy}/\sigma_0>10$. 
It has been previously shown that the physics behind shear thickening and thinning are distinct~\cite{Mari_2014, singh2018constitutive}; hence, in the following we only focus on the thickening behavior.

We observe that the extent of thickening increases with $\phi$. To characterize the steepness of the viscosity in the $\eta_{\rm} (\sigma_{xy}/\sigma_0)$ flow curve, the shear thickening
portion is fitted to $\eta_{\rm} \propto (\sigma_{xy}/\sigma_0)^\beta$, where $\beta<1$ signifies CST, $\beta = 1$ implies that the viscosity increases for unchanging shear rate
$\dot{\gamma}/\dot{\gamma}_0 = \eta_{\rm} / (\sigma_{xy}/\sigma_0)$ and hence indicates the onset of DST, and $\beta > 1$ designates DST.
In this way, we identify $\phi=0.785$ as the packing fraction at the onset of DST between two flowing states, as is evident from $\eta_{\rm r} \propto (\sigma_{xy}/\sigma_0)^\beta$ (i.e., $\beta=1$) in Fig.~\ref{fig:visc_str_model}.

The simulation data is described well by an analytical mean-field theory, which is a slight extension of that initially proposed by Wyart and Cates~\cite{wyart2014discontinuous}. This theory, used to describe the rheology and flow-state diagram of dense suspensions, centers on the fraction of frictional contacts in the system, $f({\sigma_{xy}})$, as a singular measure of the crossover between two distinct stress-independent rheologies at low and high stresses -- namely, a lubricated, frictionless branch, and a frictional branch with a non-zero value of the microscopic friction coefficient,$\mu$.

Using $f$, we can introduce stress-dependent rheological quantities that interpolate between their lubricated and fully-frictional values. In particular, the stress-dependent viscosity can be written as
\begin{equation}\label{eq:eta_str_phi}
\eta_{\rm r} (\phi,{\sigma_{xy}}) = \alpha_{\rm m}({\sigma_{xy}}) \left[\phi_{\rm m} (\sigma_{xy}) - \phi \right]^{-2},
\end{equation}
where the stress-dependent jamming volume fraction is
\begin{equation}\label{phi_str}
\phi_{\rm m}({\sigma_{xy}}) = \phi_{\rm J}^\mu  f({\sigma_{xy}}) + \phi_{\rm J}^0 [ 1-f({\sigma_{xy}}) ]~,
\end{equation}
and the stress-dependent coefficient is
\begin{equation}\label{alpha_str}
\alpha_{\rm m}({\sigma_{xy}}) =  \alpha^\mu f({\sigma_{xy}}) + {\alpha^0} (1-f({\sigma_{xy}})),
\end{equation}
in which $\alpha^{0,\mu} =  0.102, \, 0.173$ are stress-independent constants determined via fits to the viscosities of systems of frictionless and frictional particles, respectively. Finally, the fraction of frictional contacts is modeled as $f(\sigma_{xy}) = \exp \left[  -{\sigma}^{\ast}/{\sigma_{xy}}\right]$,
based on previously published results \citep{Mari_2014,Guy_2015, Ness2016,royer2016rheological, Hermes_2016, singh2018constitutive}, with $\sigma^{\ast} =0.78\sigma_0 $, as determined by fit to the $\phi=0.76$ data. 

The viscosity curves modeled by Eq.~\eqref{eq:eta_str_phi} are compared to simulation data in Fig.~\ref{fig:visc_str_model} and show good agreement, overall.

\begin{figure}[t!]
\includegraphics[width=1\linewidth]{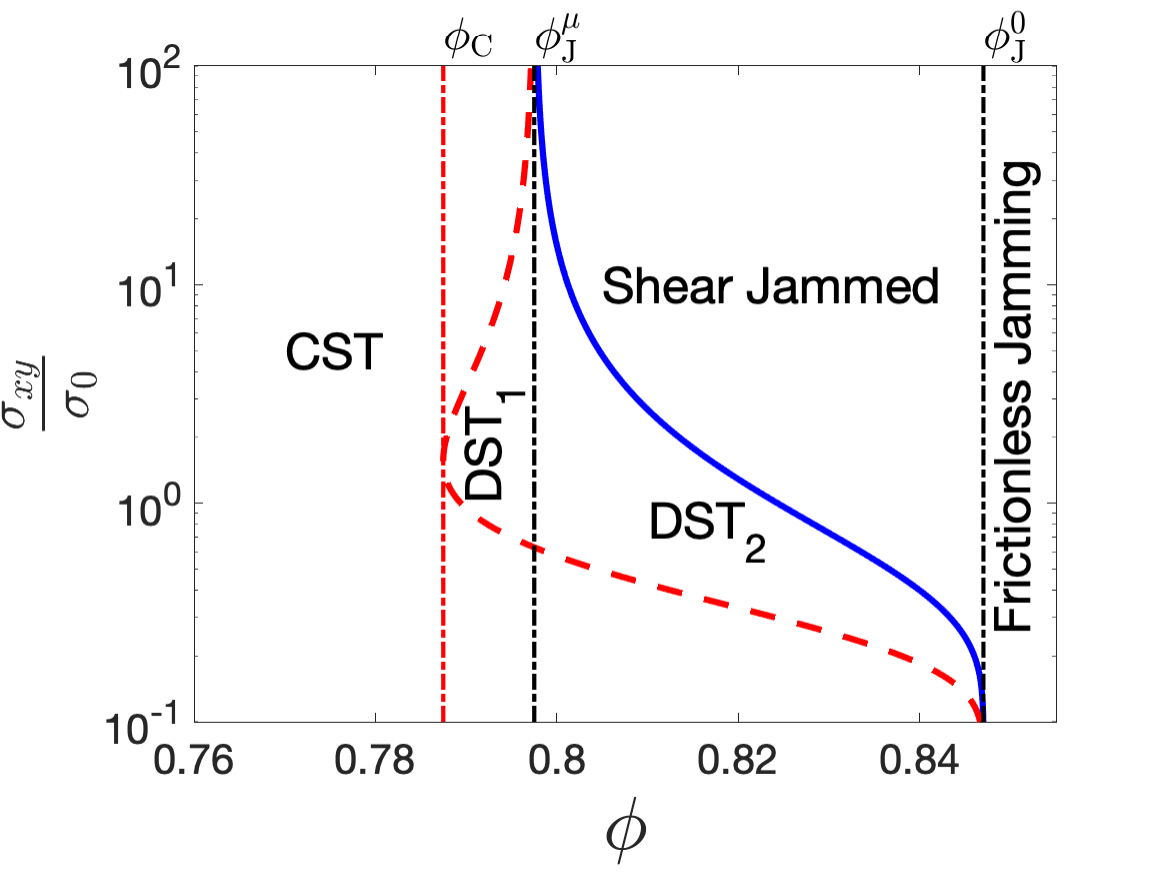}
\caption{The Phase diagram in the shear stress--packing fraction $(\sigma_{xy},\phi)$ plane. The leftmost dashed curve indicates the points where $\frac{d{\dot{\gamma}}}{d\sigma_{xy}} = 0$. The rightmost solid
curve illustrates the packing-fraction-dependent maximal stress above which the suspension is shear-jammed,
i.e., above which no flowing states exist. The two dotted-dashed vertical lines at the center and the right out of the three shown in the figure represent the frictional ($\phi_{J}^{\mu}$)
 and frictionless ($\phi_{J}^{0}$) jamming points. Finally, the dotted dashed vertical line at the left shows the minimum packing fraction $\phi_{\rm C}$ at which DST is observed.}
\label{fig:state_diagram}
\end{figure}

The fit of this mean-field model to the simulation data is presented as a flow state diagram in $\phi$ - $\sigma_{xy}/\sigma_{0}$ space in Fig.~\ref{fig:state_diagram}, in which three important packing fractions, $\phi_{\rm C}$, $\phi_{\rm J}(\mu)$, and $\phi_{\rm J}^0$, are indicated by vertical lines.
Above $\phi_{\rm J}^0$ there is no flow at any stress (without deformation of particles, which we have not explored here),
while above $\phi_{\rm J}(\mu)$ the frictional state is jammed.
$\phi_{\rm C}$ is the minimum packing fraction at which DST is observed.

In the low-stress portion of the state diagram, i.e., $\sigma_{xy}/\sigma_0 \ll 1$, the shear forces are smaller than the repulsive force, and, thus, particles do not come into contact. In this case, frictional forces are not activated, and so the rheology diverges at $\phi_{\rm J}^0$.
On the other hand, in the large-stress portion ($\sigma_{xy}/\sigma_0 \gg 1$), most of the close interactions (or \textquotedblleft contacts\textquotedblright) are frictional,
which leads to a divergence of the viscosity at $\phi_{\rm J}^\mu < \phi_{\rm J}^0$.
In these stress extremes, the viscosity is stress-independent.

For intermediate stresses ($0.3<\sigma_{xy}/\sigma_0 < 10$), continuous shear thickening is observed in the range of $\phi<\phi_{\rm C}$.
For $\phi_{\rm C} \le \phi < \phi_{\rm J}^\mu$,   DST is observed between two flowing states and is termed as DST$_1$.
The dashed line is the envelope of the DST states, with $\phi_{\rm C}$ being the point with the minimum $\phi$
value along this line. 
This line is determined as the locus of points for which $d{\dot{\gamma}}/d{\sigma_{xy}}=0$ in a flow curve $\dot{\gamma}(\sigma_{xy})$.
For $\phi> \phi_{\rm J}^\mu$, the upper boundary of the DST region is actually jammed as shown by the stress-dependent jamming line $\phi_{\rm m}({\sigma_{xy}})$,
thus DST occurs between flowing (low stress) and jammed (high stress) states, and is termed as DST$_2$.
The stress required to observe shear thickening (CST) is independent of the packing fraction, while 
the minimum stress required for DST and shear-jammed (SJ) states decreases with increase in packing fraction $\phi$. 
Eventually these curves converge and the minimum stress for jamming tends to zero as the frictionless jamming point $\phi_{\rm J}^0$ is approached.

%
%
 { The data just discussed provides one indication that the mean-field model does not capture the complete physics of the DST transition, since it does not provide a quantitative description of $\eta_r$ for $\phi=0.80$ and high stresses ($\sigma_{xy}/\sigma_0 \ge 2)$, as seen in Fig. \ref{fig:visc_str_model}.  In Sec. \rom{2} B, we show that this is the regime in which the strain-rate fluctuations exhibit significant non-Gaussian behavior.}

\subsection{\label {sec:simresults} Strain Rate Distributions} 
\label{SRDist}

{The viscosity, and flow-state diagram summarized above, provide a description of the time-averaged properties of the  DST transition that is now fairly well established~\cite{Seto_2013a,Mari_2014}.} The flow curves are obtained by computing $\langle \dot{\gamma} \rangle$,  the time average of  $\dot{\gamma}$ calculated from Eq. \ref{eq:strainrate}.
A question that has not been explored in any great depth is  how the fluctuations about the averaged quantities evolve with packing fraction and shear stress in this numerical model of DST.  Experiments indicate large-temporal fluctuations of the stress detected by a rheometer under controlled shear rate~\cite{rathee2019localized}.

\begin{figure}
\begin{subfigure}{\linewidth}
\centering
\includegraphics[scale=0.24]{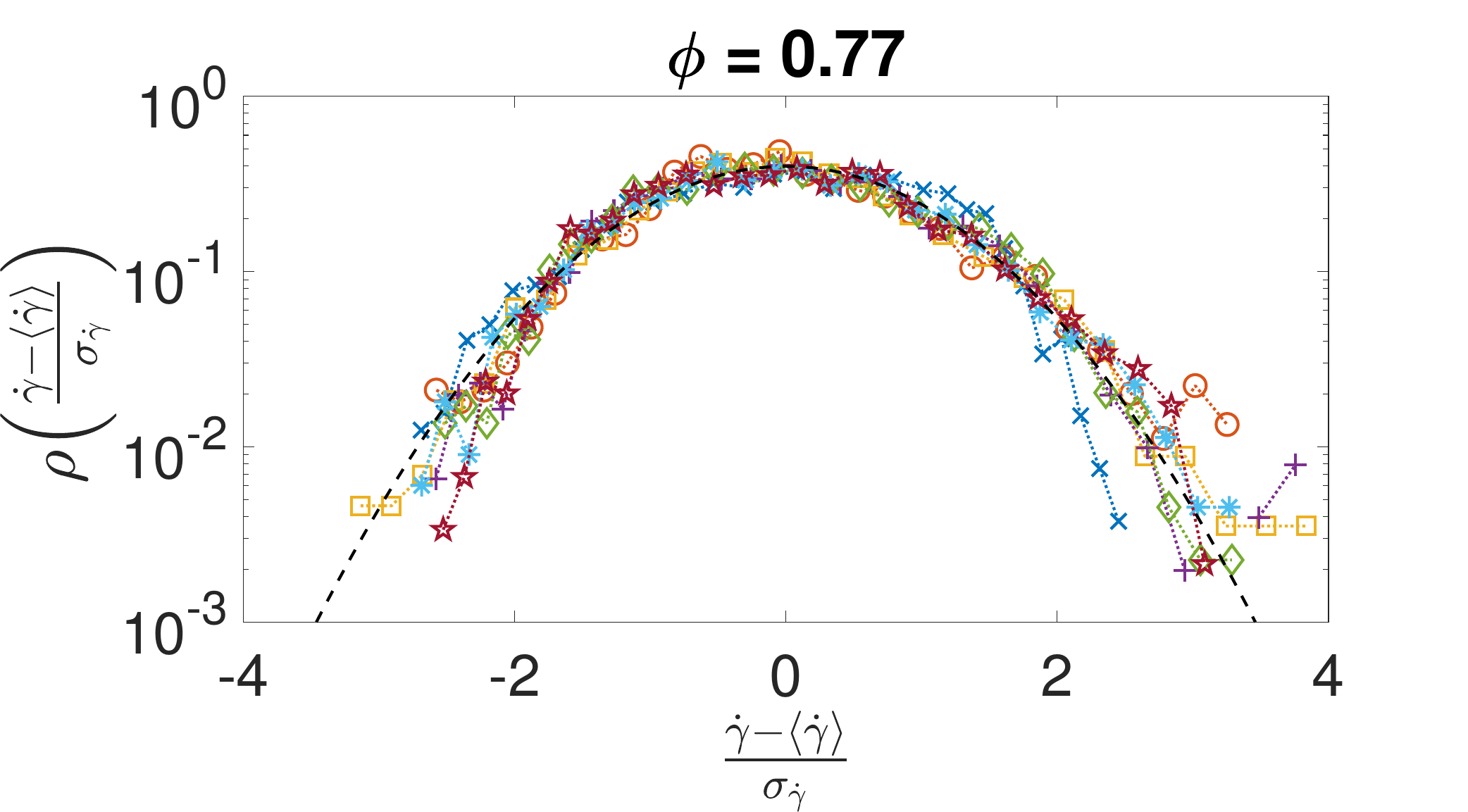}
\end{subfigure}\\[1ex]
\begin{subfigure}{\linewidth}
\centering
\includegraphics[scale=0.24]{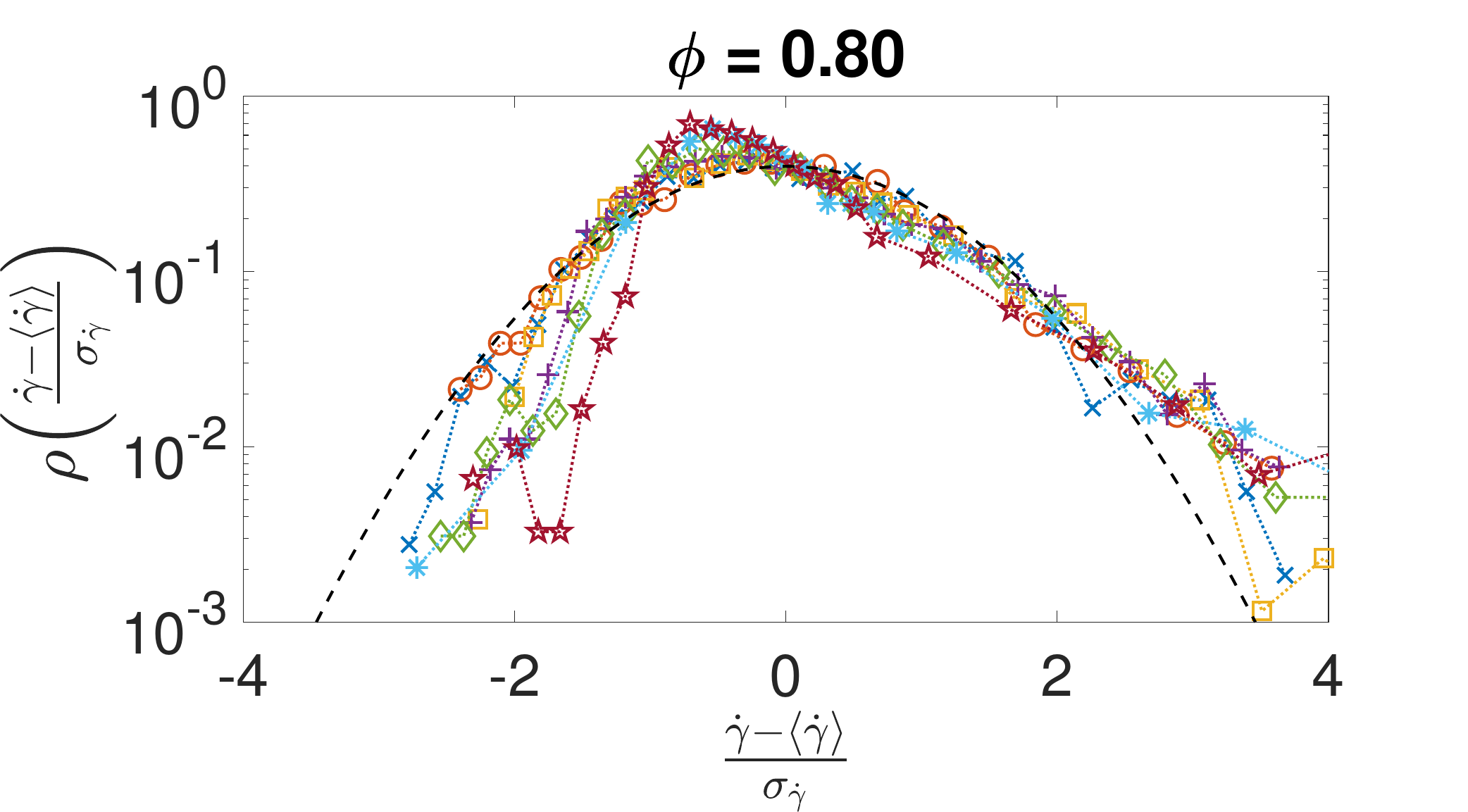}
\end{subfigure}\\[1ex]
\caption{The curves in each figure represent probability distribution of strain rates ($\dot{\gamma}$) at different stresses ($\sigma_{xy}/\sigma_{0}$) at two different  packing fractions:  $\phi = 0.77$ (top) and $\phi = 0.80$ (bottom).  The stresses at which the distributions are plotted are: $\sigma_{xy}/\sigma_{0} = 0.5$ (blue,  \ding{53}), $\sigma_{xy}/\sigma_{0} = 1.0$ (red,  \protect\markerone), $\sigma_{xy}/\sigma_{0} = 2.0$ (yellow,  \protect\markertwo), $\sigma_{xy}/\sigma_{0} = 5.0$ (purple, $+$), $\sigma_{xy}/\sigma_{0} = 10.0$ (green, \protect\markerthree), $\sigma_{xy}/\sigma_{0} = 20.0$ (cyan, \ding{83}) and $\sigma_{xy}/\sigma_{0} = 100.0$ (brown, \ding{73}). The deviations from the Gaussian (black dashed) distribution are much more pronounced at $\phi= 0.80$, as compared to $\phi = 0.77$. The average strain rate ($\langle \dot{\gamma} \rangle$) and standard deviation ($\sigma_{\dot{\gamma}}$) of the NESS have been used to scale the distributions at different stresses.}
\label{fig:strainratedistribution}
\end{figure}

In this section, we analyze the evolution of the  temporal fluctuations of $\dot{\gamma}$ as $\phi$ and $\sigma_{xy}$ are varied across the DST transition in Fig. \ref{fig:state_diagram}. 
 As seen from Fig.~\ref{fig:strainratedistribution}, the fluctuations of $\dot{\gamma}$ are distributed narrowly around the mean for $\phi$ in the CST regime ($\phi =0.77$).  In the DST regime ($\phi=0.80$), however, one observes significant non-Gaussianity in the distributions. Although the system size in the simulations is relatively small and probably not sufficient to establish the exact form of the distribution, the qualitative change we detect from $\phi =0.77$ to $\phi=0.80$ for a given system size points to the presence of longer-ranged correlations.  { It is to be noted that the non-Gaussian behavior is pronounced only within the DST region of the phase diagram where the viscosity is anomalously high, and is not a simple consequence of the $\dot \gamma =0$ cutoff on the distribution, since that would occur at any density at low enough stresses.} 

The presence of anomalous (non-Gaussian)  temporal fluctuations raises the question of what local interactions can lead to the system slowing down or speeding up as a whole.  Earlier studies of pair correlations~\cite{mari2014shear} of grains failed to identify any significant changes in positional correlations.  This observation is not inconsistent with the model of DST based on the nature of contact forces changing from lubricated to frictional, since relatively small changes in positions of the grains are all that is required to trigger this transition.  A representation that is sensitive to these changes in the contact forces are force-tiles (Maxwell-Cremona diagrams)~\cite{sarkar2013origin,2008entropytighe}.  { This was the representation previously used in constructing the statistical theory of the stress anisotropy~\cite{ref:1}.  In Sec. \rom{3}, we review the  construction of  force-tiles and the evolution of correlations in this space across the DST transition.}

 \section{\label{sec:forcetiles} Correlations in Force Space}
\label{CFS}

\begin{figure}[ht]
    \centering
    \includegraphics[scale = 0.3]{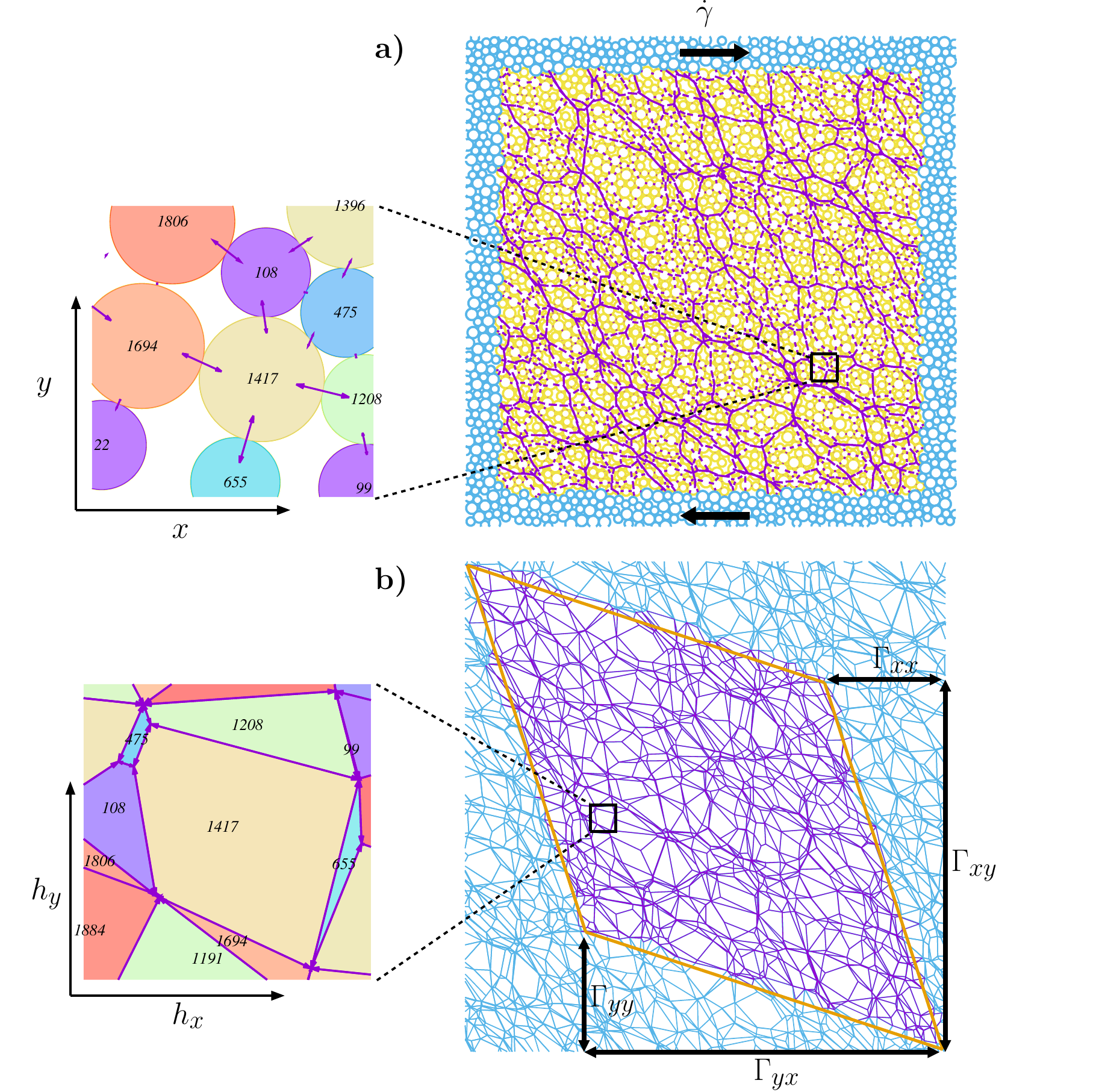}
    \caption{a) A snapshot of a suspension of 2000 soft frictional disks sheared at a variable strain rate $\dot{\gamma}$, with the shear stress $\sigma_{xy}$ held fixed~\cite{ref:1}. The lines represent the pair-wise (lubricated and frictional contact) force vectors between the individual grains. b) The force tiling associated with this flowing dense suspension. The bonds correspond to the pairwise forces, with larger polygons representing grains with higher stress. The vertices of the tiling represent height vectors $\vec{h} = (h_{x}, h_{y})$, whose difference provides the pairwise force at each bond. $\vec{\Gamma}_{x} = (\Gamma_{xx}, \Gamma_{xy})$ and $\vec{\Gamma}_{y} = (\Gamma_{yx}, \Gamma_{yy})$  represent the sum of forces in the x and y directions respectively. The regions outside the parallelogram represent periodic copies of the system.} 
    \label{fig:forcetile}
\end{figure}

In steady state, flowing suspensions provide an ensemble of microscopic states that could constitute, in principle, the basis for a statistical mechanics description of the non-equilibrium transition associated to shear thickening. The equations of motion that generate those microscopic states are determined by the constraints of force balance, as discussed in Section \ref{sec:simulation}. The idea that the forces acting on the grains and the constraints that emerge from mechanical stability can provide the right statistical ensemble to build a statistical mechanics framework for athermal jammed systems has been developed and explored in the context of granular materials (a review appears in Ref.  \cite{Bi2015}), and recently extended to shear thickening suspensions \cite{ref:1}. 
In two-dimensional systems, the force balance constraints can be naturally accounted for by working in a dual space, known as a force tiling. In a force balanced configuration of grains with pairwise forces, the “vector sum” of forces on every grain, i.e. the force vectors arranged head to tail (with a cyclic convention), form a closed polygon. Next, because of Newton’s third law, every force vector in the system has an equal and opposite counterpart that belongs to its neighboring grain. This leads to the force polygons being exactly edge-matching. Extending this to all particles within the system leads to a force tiling \cite{sarkar2013origin,2008entropytighe}. In this representation, the pairwise forces acting at each contact between two grains correspond to edges (or bonds). The vertices where the bonds meet identify vectors in this space, the vector height fields {$\vec{h}$}, such that the differences between two  such vectors connected to the same bond gives the pairwise force acting on the contact represented by that bond (see Fig. \ref{fig:forcetile}). The adjacency of the faces in the tiling is the adjacency of the grains, whereas the adjacency of the vertices is the adjacency of the voids (the heights are associated with the voids in the network). For the suspensions, in addition to the pairwise forces between grains, each particle experiences a hydrodynamic drag, which can be represented as a body force. Imposing the constraints of vectorial force balance in the presence of body forces leads to a unique solution for modified height fields, given the geometrical properties of the contact network~\cite{ramola2017stress}.

It is important to notice that the height (or force space) representation is ideally suited for exploring the statistical properties of stresses, both local and global. In the continuum, the height fields define the local Cauchy stress tensor, by the relation $\te{\sigma} = \nabla \times \vec{h}$, and the area integral of $\te{\sigma}$, or the force moment tensor, $\te{\Sigma}$ \cite{henkes2009statistical}, in terms of difference of the height fields across the system:

\begin{gather}
\te{\sigma} =
\begin{pmatrix}
\partial_{y}h_{x} & \partial_{y}h_{y}\\
-\partial_{x}h_{x} & -\partial_{x}h_{y}
\end{pmatrix};
\te{\Sigma} =
\begin{pmatrix}
L_{y}\Gamma_{yx} & L_{y}\Gamma_{yy}\\
-L_{x}\Gamma_{xx} & -L_{x}\Gamma_{xy}
\end{pmatrix}
\end{gather}
where $\vec{\Gamma}_{x(y)}$ represents the sum of forces along the $x(y)$ directions, and $L_{x(y)}$ represents the linear dimensions of the system $(\te{\sigma} = \te{\Sigma}/L_{x}L_{y})$. Additionally, global torque balance implies $\Sigma_{xy}=\Sigma_{yx}$. In our simulatons $L_{x}=L_{y}=L$, hence $\Gamma_{yy} = -\Gamma_{xx} = L\sigma_{xy}$.

At a microscopic level, each force tiling is specified by a set of vertices and a set of edges that connect these vertices. The distances between the vertices quantify the internal stress in the system, whereas the edges, which
quantify the specific contact forces in a configuration, can be thought of, in a statistical sense, as fluctuating quantities, with connections between pairs of vertices chosen with some weights. We can therefore think of the vertices of the force tilings as the points of an interacting system of particles. The effective interactions between the vertices arise from the constraints of mechanical equilibrium, and from integrating out the edges. 

Using the pair correlation function in force space (i.e. the pair correlation function of the height fields over the force tiling) one can construct an a priori probability distribution for the microscopic states (i.e. the different force tiles corresponding to the contact network of the grains under flow) and use it to define the relevant statistical ensemble to which the non-equilibrium steady states (NESS) in a dense suspension at a given $\sigma_{xy}$ and $\phi$ can be mapped~\cite{ref:1}. Such an ensemble has been proven to successfully describe the steady-state averages and the flow-state diagram. { Here we pursue a similar approach, however, we devise an explicit coarse-graining scheme to construct an effective potential that describes the interactions of height vertices separated by distances that correspond, statistically, to the non-hydrodynamic contact forces. The coarse-graining scheme relies on a separation of scales that we identify using a clustering algorithm that is presented in Sec. \rom{4}.}

 \section{\label{sec:DBSCAN} Clustering in Real Space and Force Space}
 \label{CRSFS}
 
 As we have discussed at length, the phenomenon of DST arises primarily from the switching of lubricated contacts to solid-on-solid frictional contacts triggered by changes in the positions of grains that are minuscule compared to the  size of the grains.  This separation of scales necessitates a study of correlations in both real-space and force-space.  As is visually evident from Fig. \ref{fig:forcetile},  there are correlations that exist between the positions of particles, and between the positions of the height vertices.   Two-point correlations characterized by  the pair correlations functions $g({\vec r})$ and $g^{*}({\vec h})$  provide the simplest measure of correlations.  Although $g({\vec r})$ is different from that of an ideal gas, and reflects strong excluded-volume effects, it does not evolve significantly with $\phi$ or $\sigma_{xy}$~\cite{ref:4}.  Pair correlations in force-space, $g^{*}({\vec h})$,  were used to construct an effective potential~\cite{ref:1} and a statistical model for DST.  However, the strong clustering, which creates a very large peak in $g^{*}({\vec h})$ at small $h$,  leads to a very deep minimum at $h \approx 0$ in the effective potential { because of its form ($-ln(g^{*}({\vec h}))$) that we have chosen}.  This feature hindered attempts  to carry out detailed numerical analysis of the phase behavior arising from the potential { and thereby incorporate fluctuations beyond the meanfield analysis~\cite{ref:1}. Our objective is to use the force-space approach, which provides a more sensitive measure of clustering and correlations, to construct a coarse-grained effective potential that can describe the correlations at $h$ scales that are relevant to the DST transition. To this end, we use a clustering algorithm to identify relevant scales in both real-space and $h$ space.}  
 
\begin{figure}
  \centering
  \includegraphics[scale=0.4]{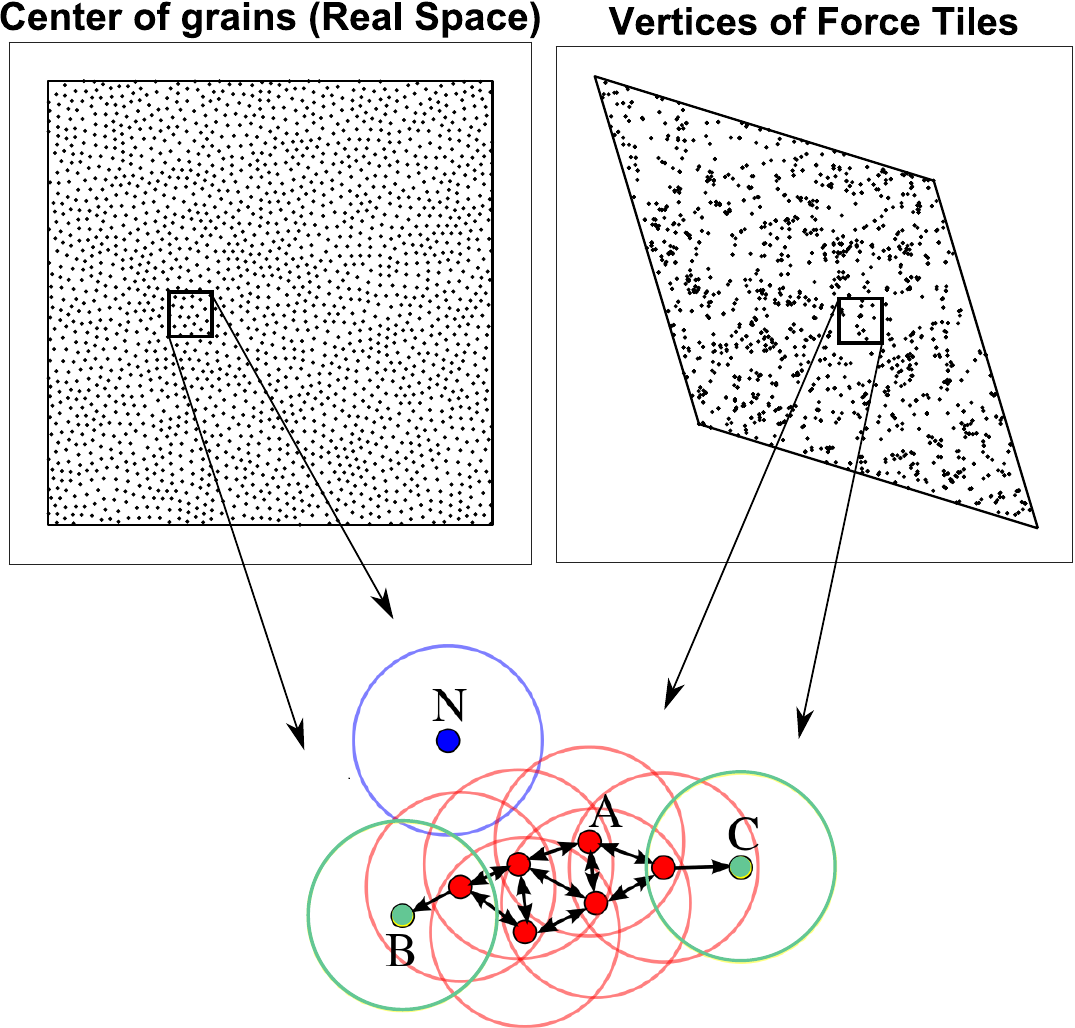}
  \caption{The top panel shows representative point patterns from the simulations in real space (centers of grains) and force-space (vertices of force tiles). The bottom panel illustrates the application of the DBSCAN algorithm \cite{ref:67} to such point patterns. The points like A, which have more than one arrow pointing towards them, belong to one cluster since all of these points can be reached by drawing circles with the probing radius $s$, centered at these points. The double-headed arrow between these points indicates that one can reach these points linked by the arrow traversing in multiple ways. The other points (B, C) are the endpoints of the cluster and can be reached only from the penultimate particle, as shown by a single-headed arrow. The point (N) is an example of a ``noise'' or an isolated point. It is the lone point of the cluster, i.e., it is not reachable from any other point by drawing circles of the radius ($s$). The \href{https://commons.wikimedia.org/wiki/File:DBSCAN-Illustration.svg\#/media/File:DBSCAN-Illustration.svg}{DBSCAN-Illustration} in the bottom panel is a slightly modified version from the original by \href{https://commons.wikimedia.org/wiki/User:Chire}{Chire} and licensed under \href{https://creativecommons.org/licenses/by-sa/3.0/deed.en}{CC BY-SA 3.0}.}
  \label{fig:DBScan_Illus}
\end{figure}

We perform a density
based clustering analysis of both grain positions and  vertices of force tiles using the DBSCAN algorithm~\cite{ref:67}.
{ In the DBSCAN technique, the set of points belonging to a single cluster consists of the union of points contained within an initial circle of probing radius $s$ centered at a given point, those contained within all circles of the same radius $s$ successively centered at all other points contained within the initial circle, and so on in an iterative sequence that continues until there are no new points contained within any subsequently-drawn circle (see Fig. \ref{fig:DBScan_Illus} for an illustration). This algorithm thereby ensures that all pairs of points from two different clusters are separated by distances greater than the probing radius $s$.}

{  In usual implementations
of DBSCAN, an optimum probing radius is determined to identify the most pronounced clustering tendencies~\cite{ref:67}.   Our aim is to use DBSCAN  to identify  characteristic clustering scales, and analyze how these evolve with packing fraction and stress.   Therefore, we do not implement the optimization procedure but instead identify the scales by measuring the number of clusters as a function of the probing radius. In addition, we do not discard any points as ``noise'' points, which is the normal practice in DBSCAN by requiring a minimum density in a cluster. In our implementation, the minimum number of points in a cluster is one such that all points in a pattern are included in some cluster.}  

The value of  $N_{c} (s)$ as $s \rightarrow 0$ has to be the total number of points in the system since each point forms its own cluster.  As $s \rightarrow {\rm system~ size}$,  $N_{c} (s) \rightarrow 1$. Our algorithm probes the density distribution
of point patterns at different length scales by monitoring
the number of clusters, $N_{c} (s)$,  as a function of the probe size $s$.   For a point pattern with uniform density,
 $N_{c} (s)$ decreases continuously with $s$. In a periodic lattice,
where the distance distribution of nearest neighbors is a
delta function,  $N_{c} (s)$ exhibits a jump discontinuity at the
lattice spacing. For a complex pattern, we expect  $N_{c} (s)$ to
show significant changes in its derivatives at  scales
where the distance distribution has structure.%

\begin{figure*}
\begin{subfigure}{.5\linewidth}
\centering
\includegraphics[scale = 0.26]{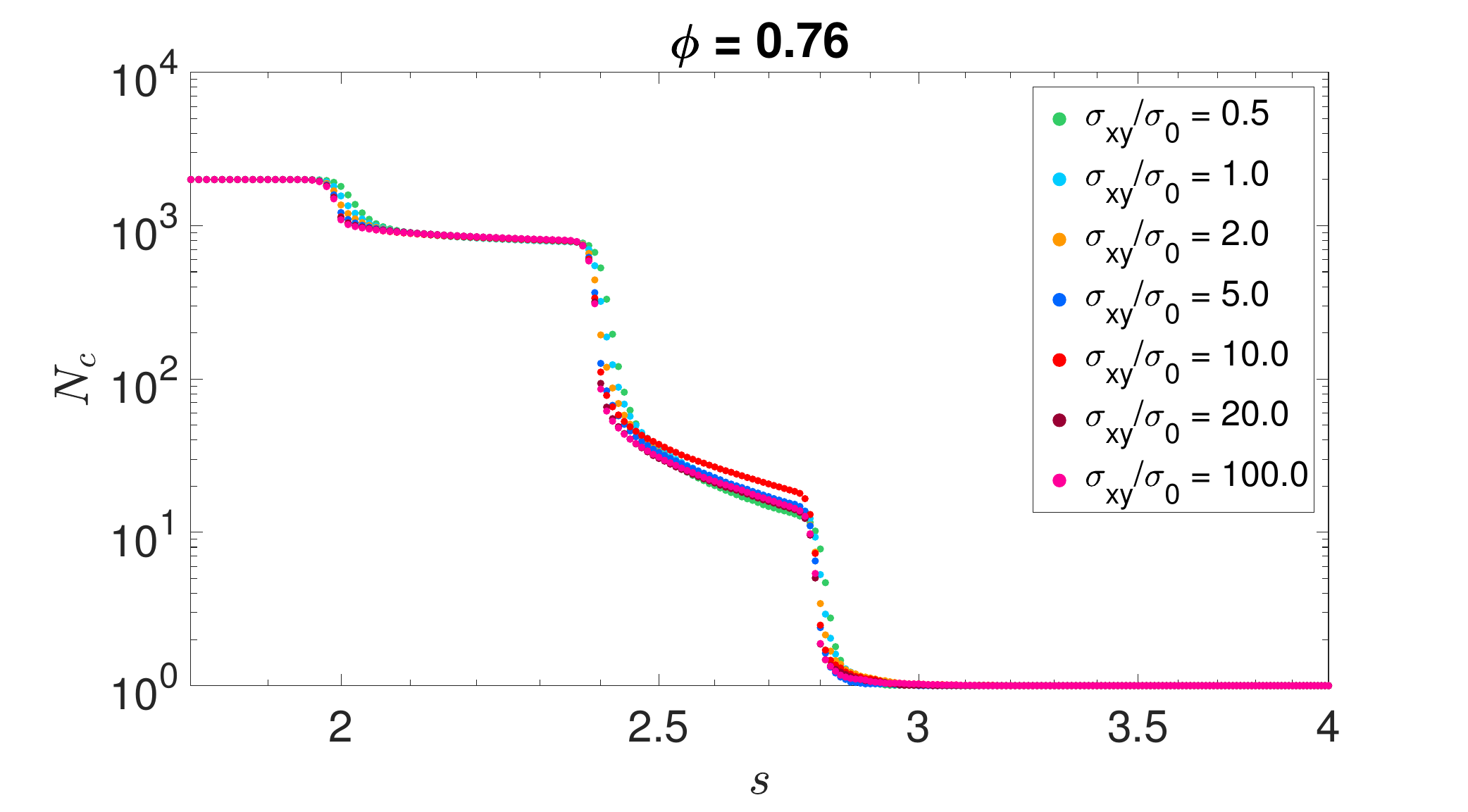}
\end{subfigure}%
\begin{subfigure}{.5\linewidth}
\centering
\includegraphics[scale = 0.26]{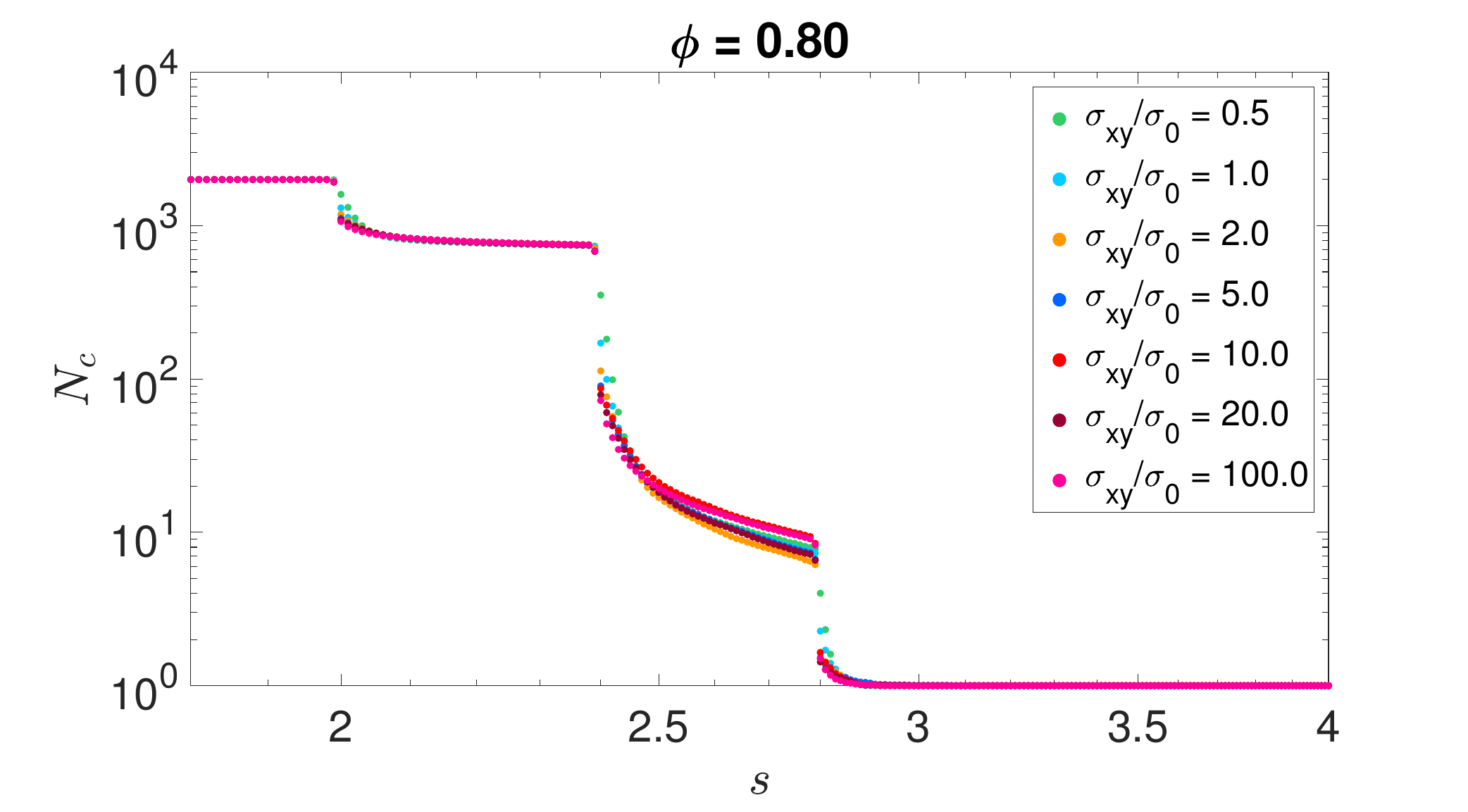}
\end{subfigure}%
\caption{Number of clusters ($N_{c}$) obtained from the DBSCAN analysis for $\phi = 0.76$ (left) and for $\phi = 0.80$ (right) for different stresses ($\sigma_{xy}/\sigma_{0}$) as function of length scale ($s$), which is measured in units of the small-grain diameter. The DBSCAN analysis is performed on the point pattern of the  centers of grains in a given configuration of the NESS.  The results are then averaged over the ensemble of configurations sampled in the dynamics. The three sharp drops in $N_{c}(s)$ at $2 \le s \le 3$ are indicative of the bidispersity of grain sizes and a clear layering of nearest neighbors.}
  \label{fig:realDBSCAN}
\end{figure*}
\begin{figure*}
\begin{subfigure}{.5\linewidth}
\centering
\includegraphics[scale = 0.26]{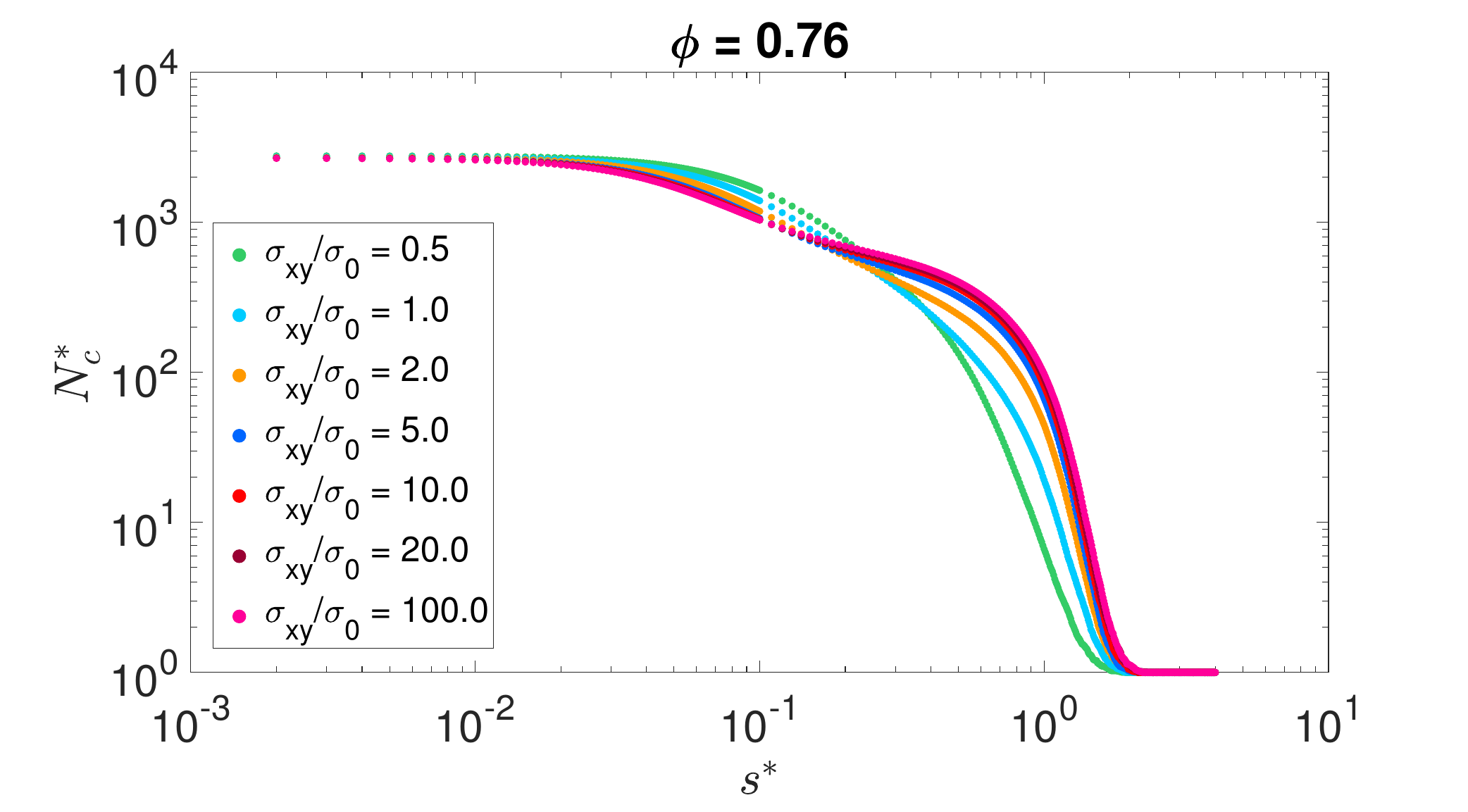}
\end{subfigure}%
\begin{subfigure}{.5\linewidth}
\centering
\includegraphics[scale = 0.26]{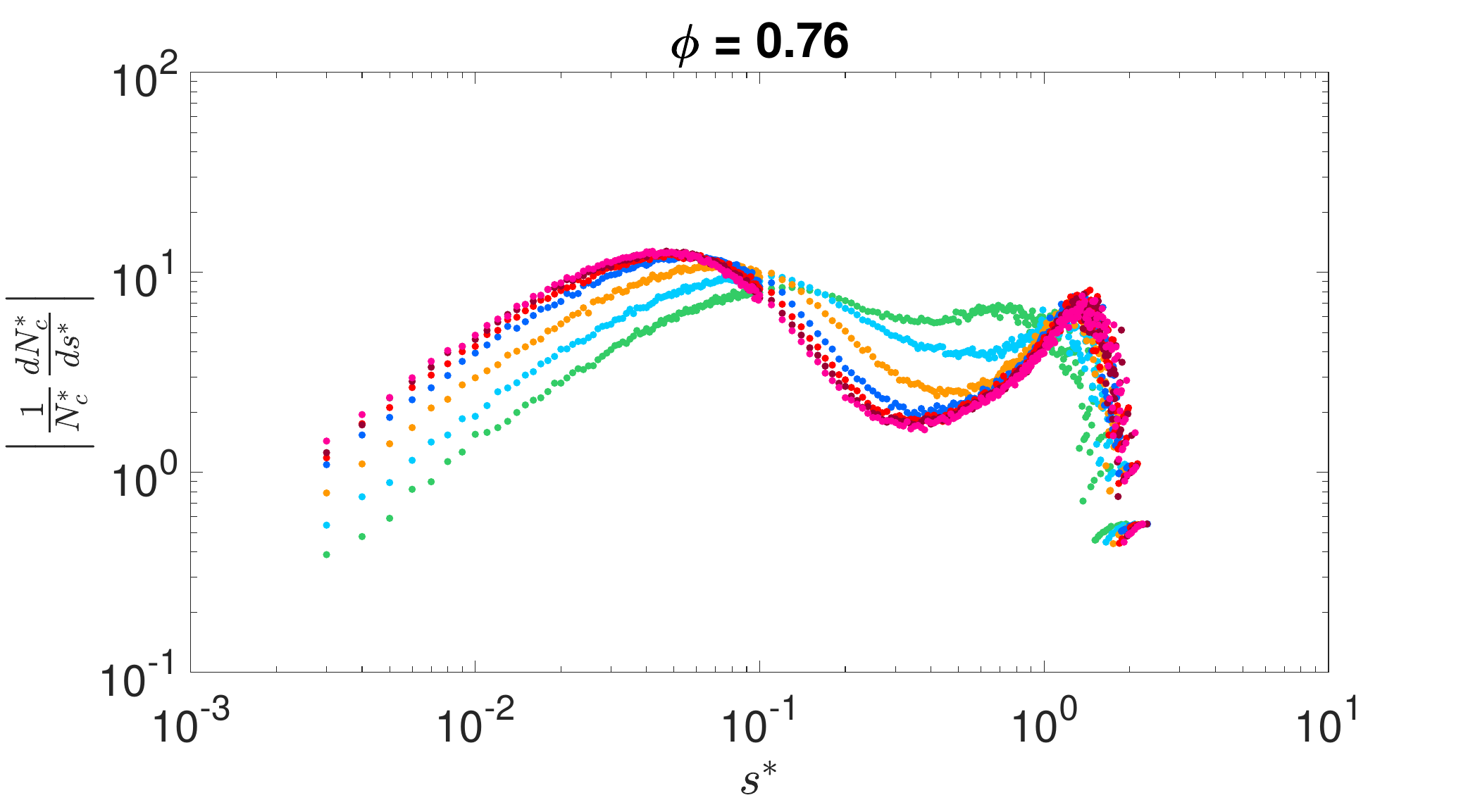}
\end{subfigure}\\[1ex]
\begin{subfigure}{0.5\linewidth}
\centering
\includegraphics[scale = 0.26]{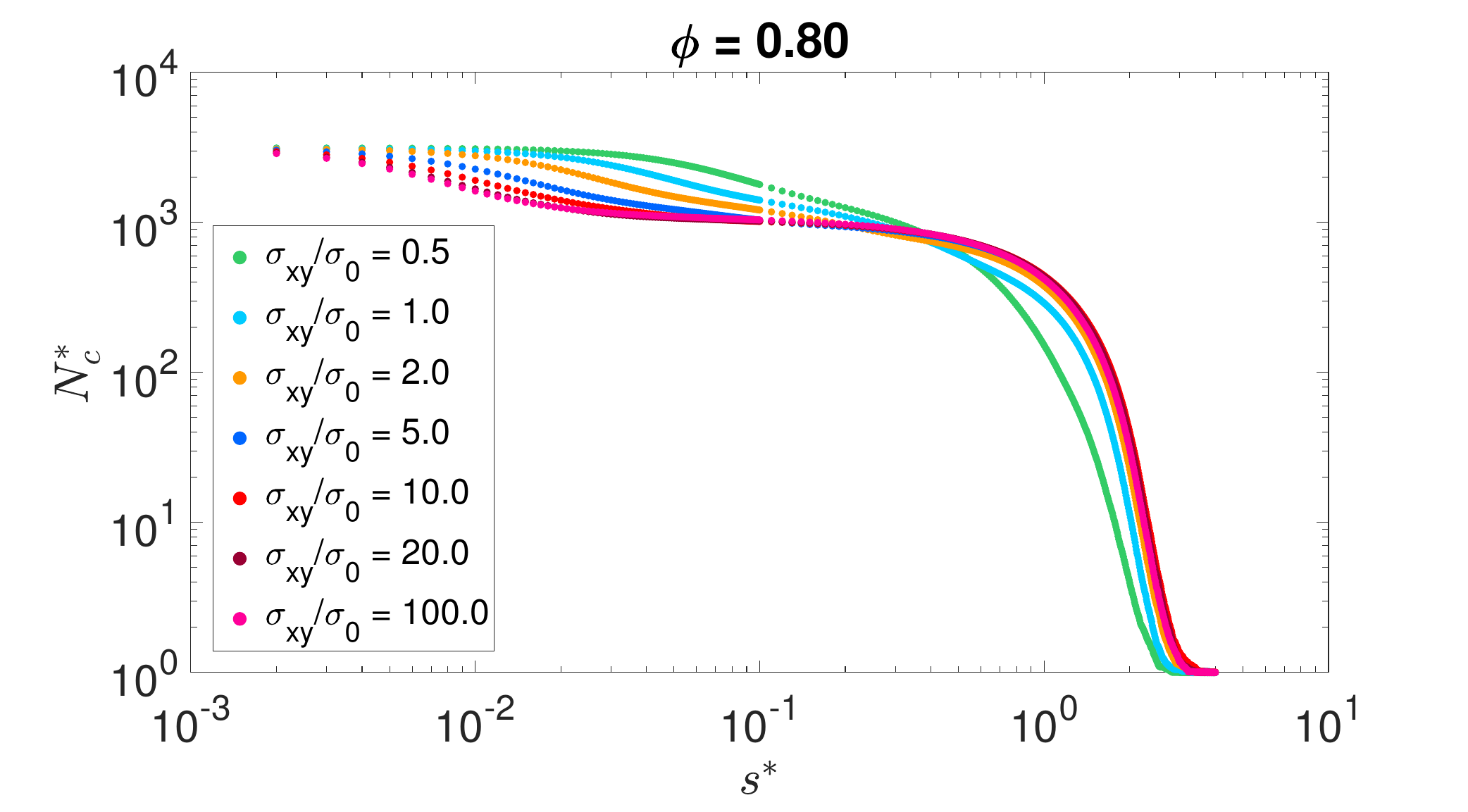}
\end{subfigure}%
\begin{subfigure}{.5\linewidth}
\centering
\includegraphics[scale = 0.26]{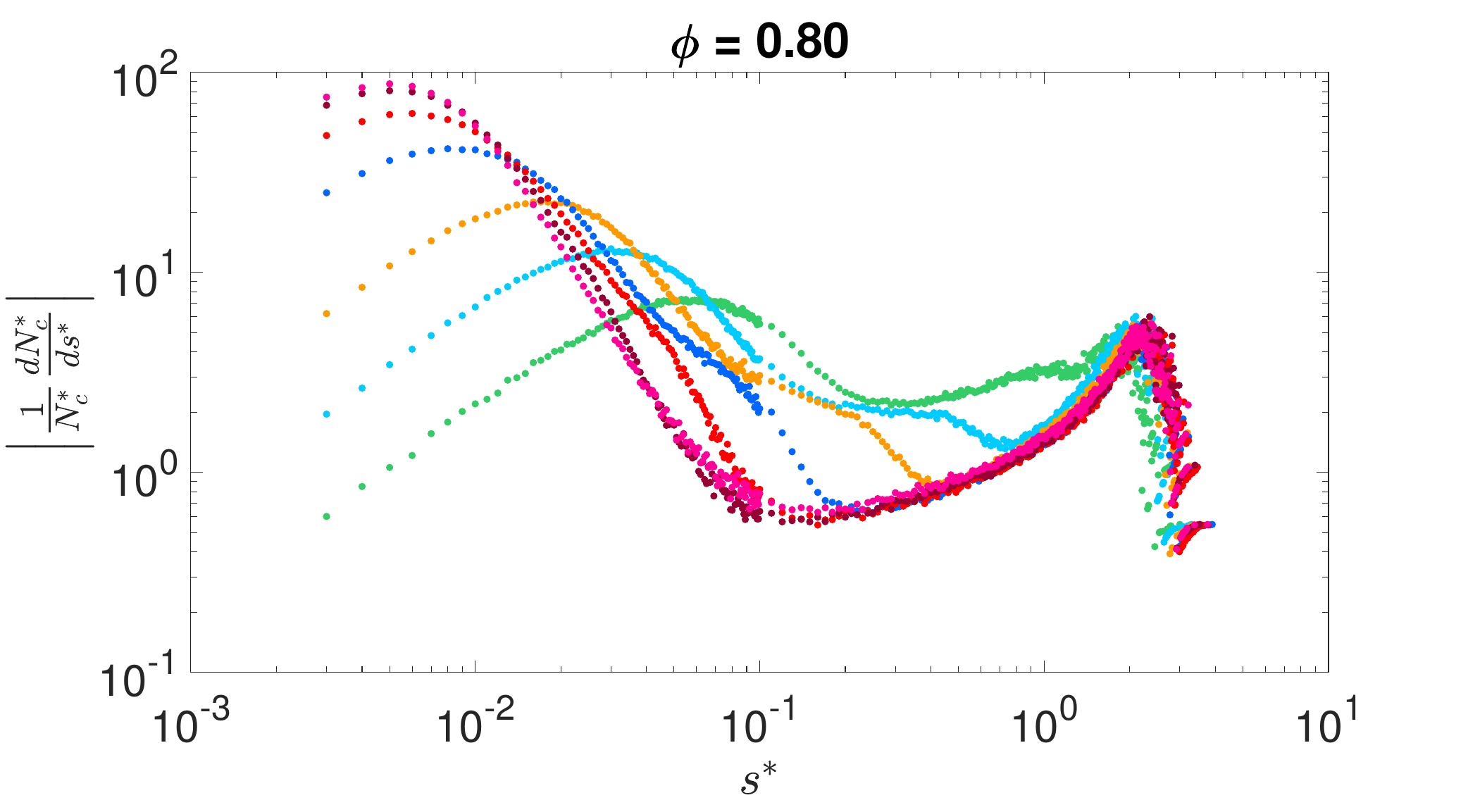}
\end{subfigure}\\[1ex]
 \caption{Number of clusters ($N^{*}_{c} \equiv N_{c}(s^{*})$) obtained from the DBSCAN analysis (left) and its derivative with respect to the probing length ($s^{*}$) (right) for different packing fractions ($\phi$) and stress ($\sigma_{xy}/\sigma_{0}$). Emergence of a  separation of scales is evident in the derivative plots. We also see in the derivative plot of DBScan at $\phi=0.80$ that (inside the DST region) a meta structure (hump in the curves) in forces emerging for $s^*<1$. The hump starts at larger $s^*\sim 1$ and goes on decreasing for larger stresses ($\sigma_{xy}/\sigma_{0}$) which isn't present outside the DST region ($\phi=0.76$).}
  \label{fig:DBSCAN}
\end{figure*}

As seen from Fig. \ref{fig:realDBSCAN},  the clustering properties of the grain centers in the CST regime ($\phi=0.76$) and the DST regime ($\phi=0.80$) are virtually identical.
The abrupt decreases in $N_c(s)$  at $s=2,2.4,2.8$ correspond to  the first layer or ordering of the nearest neighbor grains, with splitting due simply to the bidispersity. Beyond these scales, $N_c (s)$ decreases smoothly, and therefore, the density is uniform.

The situation is dramatically different for the clustering properties of height vertices (Fig. \ref{fig:DBSCAN}).  In the CST regime, $N_c(s^{*})$  decays continuously with $s^{*}$,  except for a small plateau region at intermediate values of $s^{*}$. 
In the DST regime,  we can clearly identify three different decay regimes in $N_c(s^{*})$:  an initial, relatively fast decay to a plateau with very slow decay, followed by a smooth decay.  The length of the plateau is sensitive to the stress, as is clearly evident in the plots of the derivatives.   The plateau developing at larger packing fractions and stresses is a  signature of the clustering of the small-scale clusters into meta-clusters.  The length of the plateau is a measure of the scale of this meta-clustering. It is the meta-clustering that shows significant changes across DST.
An earlier analysis~\cite{ref:15} led to similar conclusions.  We note that vertices that are close
to each other are not necessarily connected by an edge in
the tiling. Therefore, the distance distribution is not equivalent to the contact force distribution.

The structure in $N_c(s^{*})$ at small $s^{*}$ reflects the small-scale clustering of height vertices that arise, statistically,  from the small contact forces. As noted earlier, these forces arise primarily from distributing the hydrodynamic drag force between contacts, and we would like to exclude them from our statistical analysis.   Our interest is in constructing an effective potential that captures the changes brought about by switching from lubricated to frictional forces.   These occur at scales $s^{*} \ge 1$, which corresponds to forces that are comparable in magnitude  to the boundary dimensions of the force tiles (Fig. \ref{fig:forcetile}).  In Sec. \rom{5}, we present our systematic coarse-graining approach, which leads to the desired effective potential.
\section{\label{sec:potential}Constructing effective potentials}
\label{CEP}

\begin{figure*}[t]
	\centering
	\includegraphics[scale=.6]{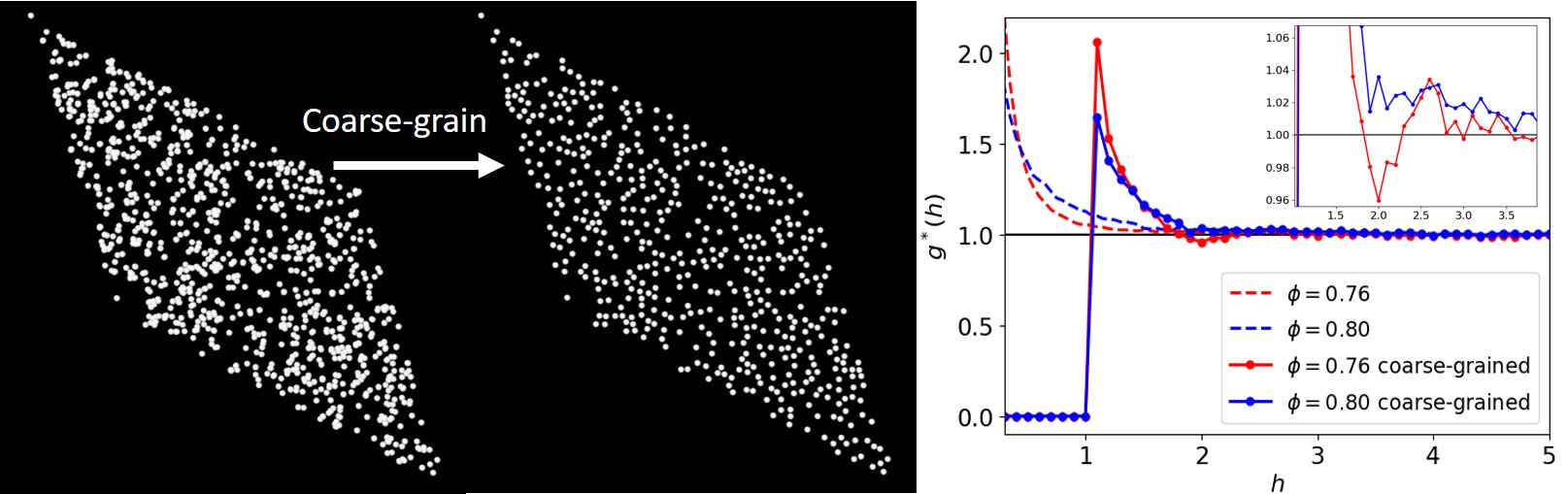}
	\caption{Snapshots show the change in our system after applying a coarse-graining procedure. While the point number has significantly decreased, the larger scale structures seem unaffected. Quantitatively,  $g^*(h)$ shows that the low $h$ peak has been cutoff at $h=1.1$ (the size of the bins we used for our clustering) and is now able to capture longer range structure that was drowned out by the extreme short-range clustering. }
	\label{fig:coarseGraining}
\end{figure*}
In order to focus on the larger forces, i.e. the larger length scale in force space as highlighted by the DBSCAN-based clustering analysis, we adopted a coarse-graining procedure for height vertices. The basic idea is to replace the tightly packed clusters of points with effective points representing these small clusters. This was carried out for each configuration by creating an empty replica, looping through the vertices randomly, and adding a point to the replica if its distance from any points already in the replica was $\geq l_{bin}$. Varying $l_{bin}$ (expressed in units of $F_0 \sigma_0 / \sigma_{xy}$, defined in Section \ref{sec:simulation}) provides different extents of coarse-graining and may change the characteristics we discuss in the following. Nevertheless, a reasonable choice is a value of $l_{bin}$ that preserves the structure in force space that is indicated by the DBSCAN analysis (Fig.\ref{fig:DBSCAN}) while removing the small-scale clustering associated primarily with hydrodynamic drag forces. In the following we discuss the results obtained for a binning size $l_{bin}=1.1$ that satisfies this criterion. In particular we compare the results obtained for a packing fraction of the grains, $\phi=0.76$, which corresponds to systems that lie in the CST part of the flow-state diagram for any value of the imposed shear stress, to those for $\phi =0.80$, which corresponds to systems that undergo DST upon increasing the imposed stress (see Fig. \ref{fig:state_diagram}).

Analyzing  the resulting coarse-grained height vertex configurations, we can see that the coarse-graining procedure decreases clumping at small distances while maintaining the same overall  distribution at larger scales (Fig. \ref{fig:coarseGraining}). Indeed, when we compute $g^*(h)$ we can clearly see the signs of longer-range structure, differently from the $g^*(h)$ computed before coarse-graining which was completely dominated by the peak near $h\approx0$. Moreover we notice that the coarse-graining procedure  reveals some qualitative differences between the pair correlations at volume fractions that correspond or not to a DST region of the flow-state diagram. In particular, compared to $\phi=0.80$, there is a dip in the correlations around $h=2$ for $\phi=0.76$. { Notably, this difference appears even in the radially-averaged $g^*(h)$ which ignores the anisotropy that was the primary focus of the earlier force-space based effective potential \cite{ref:1}. Without the coarse-graining procedure, the main difference that had been noted between volume fractions was the change in anisotropy.}
 { This observation provides an alternative approach to constructing a statistical mechanics framework based on a central potential, which facilitates numerical simulations beyond mean-field calculations.}
\begin{figure*}[t]
	\centering
	\includegraphics[scale=.7]{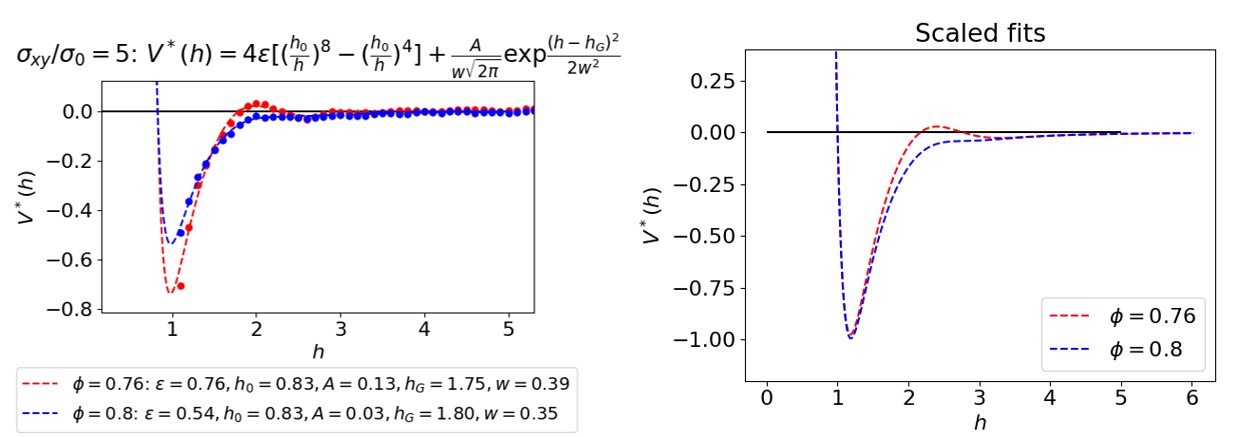}
	\caption{From the measured $g^*(h)$, it is possible to construct a pair potential that goes as $V^*(h) = -ln[g^*(h)]$. The dots in this plot represent this $V^*(h)$. To this, we fit an effective pair potential that consists of a LJ-style well plus a Gaussian bump. Fitting parameters are listed in legend. Scaling the potential by the LJ fit parameters shows the clear qualitative and quantitative differences in the shoulder at $h=2$ that depends strongly on $\phi$.}
	\label{fig:potentialFits}
\end{figure*}

We can now compute an effective pair potential from the pair correlation function of the coarse-grained height configurations as $V^*(h)=-ln[g^*(h)]$. Since by coarse-graining we have erased the correlations at $h<l_{bin}=1.1$, we avoid the problem of a very deep minimum at $h\approx0$. { Doing so reveals a short range well for both densities (i.e. both in the CST and DST regimes) and a repulsive shoulder in $V^*(h)$ around $h=2$ for the $\phi=0.76$ potential (Fig. \ref{fig:potentialFits}). This last feature appears instead to be dependent on $\phi$ and is suppressed in the DST regime---which holds true for systems across the full range of applied stresses. If we { were to construct an analogous  thermal system of particles  interacting through such a potential, the presence of a short range minimum would produce a gas-liquid phase separation with clustered non-equilibrium states in various conditions \cite{anderson-lekk}}. These clustered states are typically isotropic for a central potential. The secondary length scale introduced by the repulsive shoulder, on the other hand, can qualitatively change the equilibrium phase diagram associated with the potential and, as a consequence, also the non-equilibrium states accessible under different conditions. A competing secondary length scale can in fact introduce different spatially modulated and even anisotropic equilibrium phases which may in turn favor spatially modulated non-equilibrium states \cite{ioannidou-ncom, ciach}. 
These changes in the phase diagram can suppress, in certain cases, the gas-liquid phase separation and qualitatively change the nature of the density fluctuations \cite{Brazovskii,cates-marques}. A proper analogy between thermal systems and the force space representation of non-equilibrium steady states of shear-thickening suspensions is far from being worked out.  
However, the above observations suggest that the force space representation could be akin to a thermal system whose accessible microstates change with a change in the shape of the interaction potential, which occurs across the CST-DST boundary in sheared suspensions.}

\begin{figure*}
	\includegraphics[scale=.7]{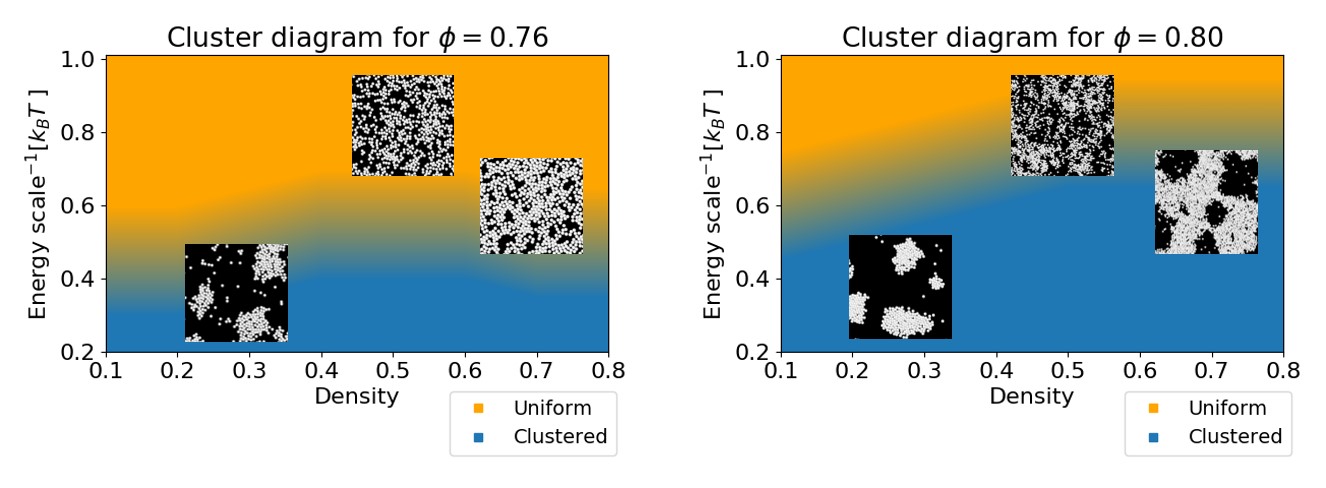}
	\caption{State diagrams of the calculated potentials corresponding to $\phi=0.80$ (top) and $\phi=0.76$ (bottom). We see a transition from a clustered to non-clustered fluid as we increase temperature. Snapshots illustrate the differences between the potentials when looking at the same temperature and density, with exactly the same simulation protocol. The types of clusters that form are different and the transition to the clustering regime happens at different temperatures.}
\label{fig:phaseBehavior}
\end{figure*}

{ To illustrate the fact that the changes in the shape of the potential detected here can modify the microstates accessible to the force space representation of the suspension, we have turned the coarse-grained effective potentials into an analytical form that can be used in molecular dynamics (MD) numerical simulations. A Lennard-Jones (LJ) style attractive well plus a Gaussian bump provides a reasonable fit of all the potentials obtained from the force space representation of the shear thickened suspensions, as shown in Fig. \ref{fig:potentialFits} (left). Scaling the fits by the LJ parameters (right) clearly shows that the repulsive bump (or shoulder) at $h\approx2$ disappears in the DST region: the scaled fits obtained for different imposed stresses lie on top of each other for $\phi=0.76$; however, they vary as a function of the stress for $\phi=0.80$, with the maximum decreasing upon crossing into the DST region. We have then performed $NVT$ MD simulations of point-like particles interacting through the coarse-grained effective potentials just discussed. \footnote{We have used the LAMMPS library to perform simulations in a 2D square box with dimensions $L_x=L_y=100$. \cite{Plimpton1995}} 

To explore the accessible states under different conditions, we vary the density and the temperature of this fictitious thermal system. The number of particles is varied between $1000$ and $8000$ while keeping the system size fixed. The particles are initially placed randomly and given a random velocity sampled from a Maxwell-Boltzmann distribution centered around the chosen temperature. The temperature used in the NVT simulations is set by the inverse energy scale, that corresponds to the maximum strength of the effective potentials \cite{frenkel-smit}. Starting with thermal energy equal to twice the potential well, we slowly reduce the temperature to the desired value. The rate is sufficiently slow that we do not see a rate dependence. Finally, we run the system at the final, fixed temperature. 

Fig.~\ref{fig:phaseBehavior} shows state diagrams constructed from the two 
potentials of Fig.~\ref{fig:potentialFits}, distinguishing clustered from non-clustered liquid states in force space, when the simulations are run in equivalent conditions, using exactly the same protocol and system size. The figure shows how clustering occurs under different conditions depending on the potential, and that the type of clustering can be qualitatively different. The effective potentials used are obtained for different densities of the suspension, and we know that varying the imposed load produces different densities of vertices in force space, but we do not yet have a clear understanding of what would play the role of temperature in the force space representation. Hence we cannot establish here a direct connection between the states sampled in the simulations (as for example shown in Fig. \ref{fig:phaseBehavior}) and those physically relevant to the force tiles of the sheared suspension.} With respect to what variable in the hydrodynamic simulations maps most closely to the temperature in the $NVT$ MD simulation of the points in force space, however, we note that the original force tiles from the hydrodynamic simulations are created from instantaneously force balanced states, which are explored under the external driving: the driving induces network rearrangements that allow the suspension to sample the microscopic states compatible with the force balance constraints. $V^*(h)$ is constructed with no information about the network reorganization, but contains information about the height vertex distribution. Is there a way to connect the sampling of the phase space associated with $V^*(h)$ via MD simulations to the way the suspension samples the microscopic NESS via real space dynamics due to the external driving? While this question requires a much deeper investigation, we speculate here that just as thermal fluctuations produce the noise that allows for phase space exploration in ordinary MD simulations, a similar role is played by the fluctuations in the shear rate of the suspensions subjected to an imposed shear stress. The fluctuations in the shear rate, such as the ones detected and discussed in Section \ref{sec:simulation}, are the manifestation of the system adjusting to the imposed driving by sampling the microstates compatible with the force balance constraints through network rearrangements. Since the network rearrangements are not explicitly included in the force space dynamics, they could translate into the noise needed to sample the force tiles associated to the real space dynamics of the grains. 

\begin{figure*}
	\includegraphics[scale=.7]{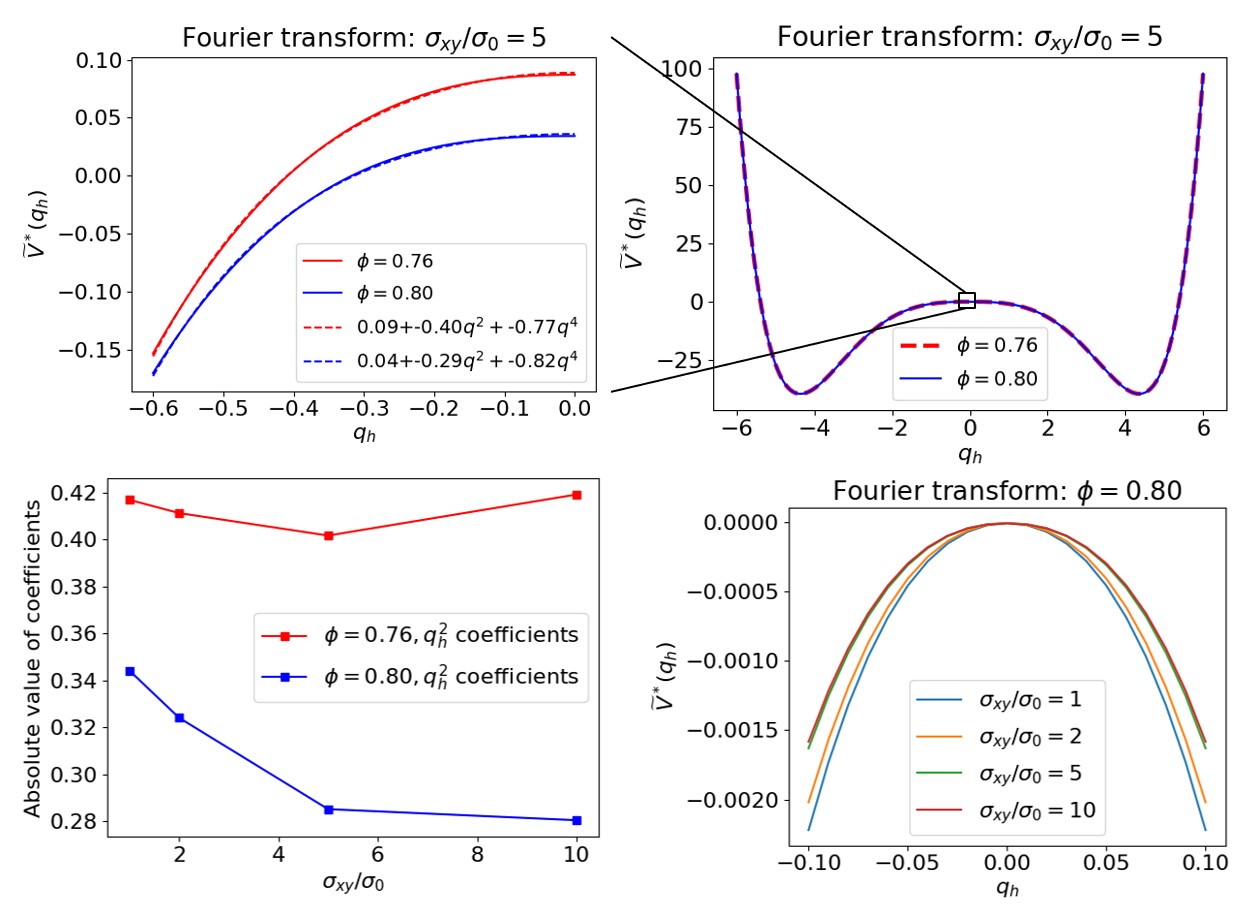}
	\caption{Fourier transforms of potentials. \textbf{Top:} The transforms deviate only at low $q_h$, which corresponds to long distances. The large $q_h$ behavior is virtually identical. At small $q_h$ the potentials can be fit by a quartic polynomial. \textbf{Bottom:} Notably, the coefficients mainly differ for the $q_h^2$ terms. When plotting these for different stress $\sigma_{xy}/\sigma_0$, we see that there is a significant drop in the DST region ($\phi=0.80$) as stress increases. The lower $q_h^2$ coefficient corresponds to a flatter $\widetilde{V}^*(q_h)$ near $q_h=0$, as seen in the bottom right panel showing the transforms for different values of $\sigma_{xy}/\sigma_0$. In this plot, a vertical shift has been applied to each transform so that $\widetilde{V}^*(0)=0$, making it easier to compare how the transforms approach this maximum for different $\sigma_{xy}/\sigma_0$.}
		\label{fig:fourier}
\end{figure*}

{ The clustering of the points in force space obtained for the two effective potentials using the MD simulations (Fig. \ref{fig:phaseBehavior}) can be mainly ascribed to their short length-scale features \cite{ioannidou-ncom}. Nevertheless,} there are intriguing difference between the effective potentials at {\it larger} length-scales that can be better highlighted by analyzing the Fourier transform (Fig. \ref{fig:fourier}). 
The low $q_h$ behavior can in general be fitted by a quartic polynomial, and plotting the $q_h^2$ coefficients as a function of stress in real space for the two volume fractions shows that the coefficients are always smaller for $\phi=0.80$ and that they do not change much with the stress at $\phi =0.76$, but they clearly decrease with the stress at $\phi=0.80$. The decrease in the $q_h^2$ coefficient, while the $q_h^4$ coefficient remains steady, must correspond to a flattening of the potential. This difference can be seen when plotting the Fourier transforms at $\phi=0.80$ for various stresses (Fig. \ref{fig:fourier}). These findings suggest that, in addition to density fluctuations associated to the minimum of $V^*(h)$, the change in the shape close to DST may promote fluctuations over larger scales. Thinking in terms of microstates that the dynamics could explore, the implications are that the shape of $\widetilde{V}^*(q_h)$ can introduce a preferred pattern in the force tiles that is characterized by the $q_h$ at which $\widetilde{V}^*$ has a minimum ($q_h^{min}$). If the $q_h^2$ coefficient is large, then a uniform configuration of contact forces will quickly reach a steady state characterized by the minimum. However, if the $q_h^2$ coefficient approaches zero, there could be long-lived transients with nearly uniform force distributions that will not resemble the structures corresponding to $q_h^{min}$. 

{ Overall, the analysis performed in this section confirms the idea that the effective potential obtained with the coarse-graining procedure, albeit radially symmetric, can still capture some of the complexity of the spatial arrangement of the vertices in force space and contains important information about the physics of the different flow-states. Moreover, intriguing similarities with the dynamics of thermal systems suggest a possible path to sample  the microscopic fluctuations of the shear thickening suspension in force space.}

\section{Conclusions}
In this paper we have sketched a possible path toward a statistical mechanics framework for shear thickening dense suspensions of grains that is based on the force space representation of the flowing suspension and naturally includes microscopic fluctuations. The overarching question is the nature of fluctuations and correlations close to DST beyond the mean field descriptions developed so far.  
{ Going beyond time-averaged properties,  we presented distributions of shear rates measured in  microscopic simulations of a numerical model~\cite{Seto_2013a,Mari_2014} of suspensions undergoing DST.  These distribution featured anomalous non-Gaussian fluctuations in the DST regime. While these fluctuations are suggestive of long-ranged microscopic correlations, microscopic measures of clustering of grain positions do not reveal any changes across the DST transition.} However, while the clustering of particles in real space remained virtually unchanged on transitioning from CST to DST, the force-tile representation of the suspension, which is based on the network of forces acting between the grains, provided further insight. In this representation, the pair-wise forces are edges.  The vertices where the edges meet define vectors in this space. The distance between the vertices quantify the internal stresses in the system. A clustering analysis, similar to the one applied in real space, revealed qualitative changes in the correlations between these height vertices as the suspension transitioned from CST to DST. 

By implementing a coarse-graining procedure, we were able to filter out the hydrodynamic drag forces and focus on the contact forces that play the dominant role in DST. From the pair correlations of these coarse-grained points, we constructed an effective pair potential. Making an analogy between force vertices and point particles interacting through an effective potential, we probed the microscopic states accessible in force space in presence of these interactions and found that the changes detected in the potential shape far away from, and close to, DST may qualitatively change the type and degree of clustering that the force tiling can undergo. By analyzing the Fourier transform of the effective potentials, we detected intriguing differences in the low $q_h$ behavior, signaling the possibility of long-ranged fluctuations. In particular, potentials constructed from force vertices in DST showed a change consistent with the presence of long-lived transients that are very different from the steady state---much like the anomalous strain rate fluctuations.

Building on this work, we will extend the MD simulations to explore the dynamical behavior more systematically.  In particular, we need to better understand how the temperature in the molecular dynamics simulations maps to network rearrangements and fluctuations in the flow of the suspension. The qualitative changes in $\tilde{V}^*(q_h)$ that accompany the CST to DST transition suggest that a fruitful avenue for going beyond mean-field theory is to construct the analog of a Ginzburg-Landau functional with the density in height space (force-tilings) serving as the order parameter~\cite{Goldenfeld}. Standard techniques can then be applied to compute correlation functions, and investigate singularities indicative of a non-equilibrium phase transition between the steady states of the suspension. 
\section{\label{sec:ack} Acknowledgements} This research was supported in part by NSF PHY11-25915. We acknowledge the hospitality of the Kavli Institute for Theoretical Physics, where part of this work was carried out. The Physics of Dense Suspensions program at KITP was instrumental in developing this collaboration. 
We acknowledge many useful discussions with Jeff Morris and Kabir Ramola.  JT, DB, and BC acknowledge the support of  NSF-CBET grants 1916877 and 1605428, and  BSF-2016118. AG and EDG acknowledge the NIST PREP Gaithersburg Program (70NANB18H151) and Georgetown University for support.  

%

\end{document}